\documentclass[12pt]{article}
\usepackage{amsmath,amssymb,amsthm,amsxtra,color,
overpic,bbm,subfigure,float,bm,epsfig}
\def\thefootnote{\fnsymbol{footnote}}
\makeatletter
\renewcommand{\@thesubfigure}{\hskip\subfiglabelskip}
\makeatother

\begin{document}

\vspace{0.2cm}

\begin{center}
{\Large\bf Indirect unitarity violation entangled with matter
effects in reactor antineutrino oscillations}
\end{center}

\vspace{0.2cm}

\begin{center}
{\bf Yu-Feng Li$^{a}$}\footnote{Email: liyufeng@ihep.ac.cn},
~ {\bf Zhi-zhong Xing$^{a, b}$}\footnote{Email: xingzz@ihep.ac.cn},
~ {\bf Jing-yu Zhu$^{a}$}\footnote{Email: zhujingyu@ihep.ac.cn} \\
{$^a$Institute of High Energy Physics, and School of Physical
Sciences, \\ University of Chinese Academy of Sciences, Beijing 100049, China \\
$^b$Center for High Energy Physics, Peking University, Beijing
100080, China}
\end{center}

\vspace{1.5cm}

\begin{abstract}
If finite but tiny masses of the three active neutrinos are generated via
the canonical seesaw mechanism with three heavy sterile neutrinos,
the $3\times 3$ Pontecorvo-Maki-Nakagawa-Sakata neutrino mixing
matrix $V$ will not be exactly unitary. This kind of indirect
unitarity violation can be probed in a precision reactor
antineutrino oscillation experiment, but it may be entangled
with terrestrial matter effects as both of them are very small.
We calculate the probability of $\overline{\nu}^{}_e \to
\overline{\nu}^{}_e$ oscillations in a good analytical approximation,
and find that, besides the zero-distance effect, the effect
of unitarity violation is always smaller than matter effects,
and their entanglement does not appear until the next-to-leading-order
oscillating terms are taken into account. Given a 20-kiloton JUNO-like
liquid scintillator detector, we reaffirm
that terrestrial matter effects should not be neglected but
indirect unitarity violation makes no difference,
and demonstrate that the experimental sensitivities to the neutrino
mass ordering and a precision measurement of $\theta^{}_{12}$ and
$\Delta^{}_{21} \equiv m^2_2 - m^2_1$ are robust.
\end{abstract}

\begin{flushleft}
\hspace{0.8cm} PACS number(s): 14.60.Pq, 14.60.St, 13.15.+g.
\end{flushleft}

\def\thefootnote{\arabic{footnote}}
\setcounter{footnote}{0}

\newpage

\section{Introduction}

Experimental neutrino physics is entering the era of precision
measurements, in which some fundamental questions about the properties
of massive neutrinos will hopefully be answered. One of the burning
issues is whether there exist some extra (sterile) neutrino species
which do not directly participate in the standard weak interactions.
Such hypothetical neutrinos are well motivated in
the canonical (type-I) seesaw mechanism
\cite{Fritzsch:1975sr,Minkowski:1977sc,Sawada:1979dis,
VanNieuwenhuizen:1979hm,Levy:1980ws,Mohapatra:1979ia,
Schechter:1980gr},
which works at a
high energy scale far above the electroweak symmetry breaking scale ---
it can naturally generate finite but tiny Majorana masses for the
standard-model neutrinos (i.e., the mass eigenstates $\nu^{}_1$,
$\nu^{}_2$ and $\nu^{}_3$ corresponding to the flavor eigenstates
$\nu^{}_e$, $\nu^{}_\mu$ and $\nu^{}_\tau$) and interpret the
observed matter-antimatter asymmetry of the Universe via the
canonical leptogenesis mechanism \cite{FY}
\footnote{Throughout this work we only focus on the seesaw-induced
heavy sterile neutrinos. Light sterile neutrinos have also attracted
a lot of phenomenological attention, but in general they are less
motivated from a theoretical point of view, although they have been
introduced so as to explain some ambiguous ``anomalies" \cite{Sterile,Gariazzo:2015rra,Gariazzo:2017fdh,Dentler:2017tkw,Fong:2017gke}.}.
Assuming the existence
of three heavy sterile neutrinos in this seesaw picture,
one may write out the standard weak charged-current interactions in
terms of the mass eigenstates of three charged leptons and six neutrinos
as follows:
\begin{eqnarray}
-{\cal L}^{}_{\rm cc} = \frac{g}{\sqrt{2}} \overline{\begin{pmatrix}
e & \mu & \tau\end{pmatrix}^{}_{\rm L}} ~ \gamma^\mu \left[
V \begin{pmatrix} \nu^{}_1 \cr \nu^{}_2 \cr \nu^{}_3 \cr
\end{pmatrix}_{\hspace{-0.15cm} \rm L}
+ R \begin{pmatrix} \nu^{}_4 \cr \nu^{}_5 \cr \nu^{}_6 \cr
\end{pmatrix}_{\hspace{-0.15cm} \rm L}
\right] W^-_\mu + {\rm H.c.} \; ,
\end{eqnarray}
where $\nu^{}_4$, $\nu^{}_5$ and $\nu^{}_6$ stand for the three
heavy-neutrino mass eigenstates, $V$ is the $3\times 3$
Pontecorvo-Maki-Nakagawa-Sakata (PMNS) flavor mixing matrix \cite{P,MNS},
and $R$ is a $3\times 3$ matrix describing the strength of flavor
mixing between $(e, \mu, \tau)$ and $(\nu^{}_4, \nu^{}_5, \nu^{}_6)$.
Because $V V^\dagger = {\bf 1} - R R^\dagger$ holds \cite{Xing2007}, where
${\bf 1}$ denotes the identity matrix, the PMNS matrix $V$ is not
exactly unitary. Following the full angle-phase parametrization of the whole
$6\times 6$ neutrino mixing matrix advocated in Refs.
\cite{Xing:2009vb, Xing2011} and
taking account of the fact that all the mixing angles appearing in $R$
must be very small, it is convenient to express $V$ as
$V = \left({\bf 1} - \kappa\right) U$, in which
\begin{eqnarray}
&& U = \begin{pmatrix}
U^{}_{e1} &  U^{}_{e2} & U^{}_{e3} \cr
U^{}_{\mu 1} &  U^{}_{\mu 2} & U^{}_{\mu 3} \cr
U^{}_{\tau 1} &  U^{}_{\tau 2} & U^{}_{\tau 3} \cr \end{pmatrix}
= \begin{pmatrix} c^{}_{12} c^{}_{13} & {s}^{}_{12} c^{}_{13} &
\hat{s}^*_{13} \cr -s^{}_{12} c^{}_{23} - c^{}_{12} \hat{s}^{}_{13}
s^{}_{23} & c^{}_{12} c^{}_{23} - s^{}_{12} \hat{s}^{}_{13}
s^{}_{23} & c^{}_{13} s^{}_{23} \cr s^{}_{12} s^{}_{23}
- c^{}_{12} \hat{s}^{}_{13} c^{}_{23} & -c^{}_{12} s^{}_{23}
-s^{}_{12} \hat{s}^{}_{13} c^{}_{23} & c^{}_{13} c^{}_{23} \cr\end{pmatrix}
\; , \hspace{1.5cm}
\nonumber \\
&& \kappa \simeq \frac{1}{2} \begin{pmatrix} \kappa^{}_{11} & 0 & 0 \cr
0 & \kappa^{}_{22} & 0 \cr 0 & 0 & \kappa^{}_{33} \cr \end{pmatrix} +
\begin{pmatrix} 0 & 0 & 0 \cr \kappa^{}_{21} & 0 & 0 \cr
\kappa^{}_{31} & \kappa^{}_{32} & 0 \cr \end{pmatrix}
\end{eqnarray}
with $\kappa^{}_{ij} = \hat{s}^{*}_{i 4} \hat{s}^{}_{j 4} +
\hat{s}^{*}_{i 5} \hat{s}^{}_{j 5} + \hat{s}^{*}_{i 6} \hat{s}^{}_{j 6}$
for $i \geq j = 1, 2, 3$. Here the notations $c^{}_{ij} \equiv \cos\theta^{}_{ij}$,
$s^{}_{ij} \equiv \sin\theta^{}_{ij}$ and $\hat{s}^{}_{ij} \equiv s^{}_{ij}
e^{{\rm i}\delta^{}_{ij}}$ have been used, where $\theta^{}_{ij}$ and
$\delta^{}_{ij}$ are the rotation and phase angles, respectively.
It is obvious that nonzero $\kappa^{}_{ij}$ arise from the small
mixing between light
and heavy neutrino states described by $\theta^{}_{ij}$ (for $i=1,2,3$
and $j=4,5,6$), and therefore they measure the deviation of $V$
from $U$ --- the effect of {\it indirect} unitarity violation (UV)
caused by the heavy degrees of freedom which do not directly take part
in the low-energy lepton-flavor-violating processes, such as neutrino
oscillations.
The current limits on the {\it indirect} UV effect can be found in Refs.~\cite{Antusch2006,Antusch:2014woa,Fernandez-Martinez:2016lgt,
Blennow:2016jkn}, where the elements of $|VV^{\dagger}|=|(1-\kappa) (1-\kappa^{\dagger})|$ are constrained from the electroweak precision observables,
low energy weak measurements and the neutrino oscillation data. A typical and conservative expectation is that the magnitude of $\kappa_{ij}^{}$ should
be smaller than 0.05, which indicates that the active-sterile mixing angles $\theta^{}_{ij}$ (for $i=1,2,3$
and $j=4,5,6$) can be taken as large as $7.5^{\circ}$.
So far a lot of attention has been paid to possible effects of indirect UV
in the accelerator-based long-baseline neutrino oscillation experiments
\cite{Antusch2006,Antusch:2009pm,Escrihuela:2015wra,Bekman:2002zk,Ge:2016xya,Miranda:2016wdr,Escrihuela:2016ube,C:2017scx,Tang},
and limited attention has also been given to this kind of effect in a
reactor-based antineutrino oscillation experiment \cite{Xing2013,Qian,Fong2016}.
It is already known that the UV-induced ``zero-distance effect" must appear
in the ``disappearance" oscillation probability $P(\overline{\nu}^{}_\alpha \to
\overline{\nu}^{}_\alpha)$ (for $\alpha = e, \mu, \tau$) \cite{Xing2013,Langacker:1988up},
for example,
\begin{eqnarray}
P(\overline{\nu}^{}_e \to \overline{\nu}^{}_e)|^{}_{L =0} = c^4_{14}
c^4_{15} c^4_{16} \simeq 1 - 2 \kappa^{}_{11} \; ,
\end{eqnarray}
but extracting this small effect is extremely difficult even though there
is a near detector, because uncertainties associated with the reactor
antineutrino flux are expected to be overwhelming considering the
reactor antineutrino anomaly and spectral features for the reactor
antineutrino fluxes at around 5 MeV. In this case one may
wonder whether the oscillating terms of $P(\overline{\nu}^{}_e \to
\overline{\nu}^{}_e)$ can provide some information about the indirect UV
or not
\footnote{In the subsequent analytical calculations and numerical
simulations we shall only focus the UV effect in the oscillating
terms, and neglect the UV-induced zero-distance
effect. The latter has been discussed, for example, in
Ref.~\cite{Xing2013} and Chapter 3 of Ref.~\cite{JUNO}.}.

As pointed out in Refs. \cite{Capozzi:2013psa,Capozzi:2015bpa,Wang}, terrestrial matter effects should not
be neglected in the JUNO-like reactor antineutrino oscillation experiment with
the baseline length $L \simeq 53 ~{\rm km}$ \cite{JUNO},
since their strength is essentially
comparable with the experimental sensitivity to the neutrino mass
ordering. Two natural and meaningful questions 
turn out to be:
(a) how the indirect UV effect is entangled with matter effects in
$\overline{\nu}^{}_e \to \overline{\nu}^{}_e$ oscillations; (b)
whether they can be distinguished from each other. The main purpose of
the present work is just to answer these two questions.

The remaining parts of this paper are organized as follows. In section 2
we derive the analytical expression of
$P(\overline\nu_{e}^{} \to \overline\nu_e^{})$
by including both indirect UV and terrestrial matter effects
and making a good approximation for the antineutrino
beam energy of a few MeV \cite{Zhu, Li:2016pzm}.
Section 3 is devoted to some numerical simulations based on
the setup of a JUNO-like detector in order to answer the
above two questions. We find that the indirect UV effect is always smaller
than terrestrial matter effects, and their entanglement does not
appear until the next-to-leading-order oscillating terms
are taken into account. We summarize our main results in section 4
with two concluding remarks: (a) indirect UV makes no difference in
the JUNO-like experiment; (b) such an experiment's sensitivities to the
neutrino mass ordering and a precision measurement of $\theta^{}_{12}$ and
$\Delta^{}_{21} \equiv m^2_2 - m^2_1$ are robust.

\section{Analytical approximations of
${P}(\overline \nu^{}_e \to \overline\nu^{}_e)$}

Of course, the three heavy sterile neutrinos are kinematically
forbidden to take part in neutrino oscillations in any realistic
accelerator- or reactor-based experiments.
Given the indirect UV effect hidden in the PMNS matrix $V$,
the effective Hamiltonian describing the propagation of the {\it
antineutrino mass eigenstates} in matter with a constant
density profile can be written as
\begin{eqnarray}
\widetilde {\cal H} = \begin{pmatrix}
E_1^{} & 0 & 0 \cr 0 & E_2^{} & 0 \cr 0 & 0 & E_3^{} \cr\end{pmatrix}
- \frac{G_{\rm F}^{}}{\sqrt{2}} V^{\rm T}
\begin{pmatrix} 2 N^{}_e - N^{}_n & 0 & 0 \cr
0 & -N^{}_n & 0 \cr 0 & 0 & -N^{}_n \cr\end{pmatrix} V^{*} \; ,
\end{eqnarray}
where $E_i^{} \simeq E + m^{2}_i/\left(2 E\right)$
with $E$ and
$m_i^{}$ being the beam energy and masses of antineutrinos
respectively (for $i=1,2,3$),
$G_{\rm F}^{}$ denotes the Fermi constant, $N_e^{}$ and $N_n^{}$
stand respectively for the electron and neutron densities in matter.
It is clear that the neutral-current-induced coherent forward
scattering effect (described by $N^{}_n$) becomes trivial and
negligible, if $V$ is exactly unitary. Now this effect, together
with the charged-current-induced coherent forward scattering effect
(described by $N^{}_e$ and only sensitive to the $e$-flavored neutrinos
and antineutrinos), constitutes the terrestrial matter effect
and can thus modify the behavior of antineutrino oscillations.
{{Note that in Eq.~(4) and throughout this paper we denote all the quantities in matter with tilde hats
as their counterparts of the corresponding vacuum quantities in the indirect UV framework.}}

We begin with the useful formula of the
matter-modified antineutrino oscillation probability
$\widetilde P (\overline\nu^{}_{e} \to \overline\nu^{}_{e})$
derived by Kimura, Takamura and Yokomura (KTY)
\cite{Kimura:2002wd,Yasuda:2007jp} and take account of
the indirect UV effect \cite{FernandezMartinez:2007ms}:
\begin{eqnarray}
\widetilde P (\overline\nu^{}_{e} \to \overline\nu^{}_{e})
\hspace{-0.17cm}& = &\hspace{-0.17cm}
\frac{1}{\left(V V^{\dag}\right)^{2}_{ee}}\left[\left|\left(V^{*}V^{\rm T}
\right)_{ee}^{}\right|^2
- 4 \sum_{j<k}^{} {\rm Re} \left(\widetilde{X}_j^{ee}
\widetilde{X}_k^{ee*}\right) \sin^2 \left(\frac{\Delta
\widetilde{E}_{jk}^{} L}{2}\right) \right] \; ,
\end{eqnarray}
where $\Delta \widetilde{E}^{}_{jk} \equiv \widetilde{E}^{}_j -
\widetilde{E}^{}_k $, $L$ denotes the baseline length
and $\widetilde{X}_j^{ee} \equiv \left(V^{*} W\right)^{}_{e j}
\left(V W^*\right)^{}_{e j}$ (for $j,k =1, 2, 3$) with
$\widetilde E^{}_i$ being the eigenvalues of $\widetilde{\cal H}$
and $W^{}_{ij}$ being the unitary matrix which diagonalizes
$\widetilde{\cal H}$ (i.e., $W^{\dag} \widetilde{\cal H} W =
{\rm Diag} \{\widetilde E^{}_1, \widetilde E^{}_2, \widetilde E^{}_3\}$).
To be explicit,
\begin{eqnarray}
\widetilde X_{j}^{ee} =
\sum^{3}_{k=1} N^{}_{jk} Y_k^{ee} \; ,
\end{eqnarray}
in which
\begin{eqnarray}
N = \begin{pmatrix}
\displaystyle \frac{\widetilde{E}^{}_{2} \widetilde{E}^{}_{3}}
{\Delta \widetilde{E}^{}_{21}
\Delta \widetilde{E}^{}_{31}} & \displaystyle -\frac{\widetilde{E}^{}_{2}
+\widetilde{E}^{}_{3}}{\Delta \widetilde{E}^{}_{21}
\Delta \widetilde{E}^{}_{31}}& \displaystyle
\frac{1}{\Delta \widetilde{E}^{}_{21}
\Delta \widetilde{E}^{}_{31}} \cr \vspace{-0.4cm} \cr
\displaystyle -\frac{\widetilde{E}^{}_{1}
\widetilde{E}^{}_{3}}{\Delta \widetilde{E}^{}_{21}
\Delta \widetilde{E}^{}_{32}} & \displaystyle \frac{\widetilde{E}^{}_{1}
+\widetilde{E}^{}_{3}}{\Delta \widetilde{E}^{}_{21}
\Delta \widetilde{E}^{}_{32}}& \displaystyle -
\frac{1}{\Delta \widetilde{E}^{}_{21}
\Delta \widetilde{E}^{}_{32}} \cr \vspace{-0.45cm} \cr
\displaystyle \frac{\widetilde{E}^{}_{1}
\widetilde{E}^{}_{2}}{\Delta \widetilde{E}^{}_{31}
\Delta \widetilde{E}^{}_{32}} & \displaystyle -\frac{\widetilde{E}^{}_{1}
+\widetilde{E}^{}_{2}}{\Delta \widetilde{E}^{}_{31}
\Delta \widetilde{E}^{}_{32}} & \displaystyle
\frac{1}{\Delta \widetilde{E}^{}_{31}
\Delta \widetilde{E}^{}_{32}} \cr\end{pmatrix} \; ,
\end{eqnarray}
and $Y_k^{ee}=(V^{*} \widetilde {\cal H}^{k-1}V^{T})^{ee}$.
Since $\widetilde X^{ee}_j$ are real and
$\Delta \widetilde E_{ij}^{}=\widetilde\Delta_{ij}^{}/2E$ with
$\widetilde \Delta_{ij}^{} \equiv \widetilde{m}^2_i - \widetilde{m}^2_j$,
the expression of $\widetilde P (\overline\nu_{e}^{} \to \overline\nu^{}_{e})$
in Eq. (5) can be rewritten as:
\begin{eqnarray}
\widetilde P (\overline\nu^{}_{e} \to \overline\nu^{}_{e})
\hspace{-0.17cm}& = &\hspace{-0.17cm}  1
- 4 \widehat{X}_1^{ee} \widehat{X}_2^{ee} \sin^2 \widetilde F_{21}^{}
- 4 \widehat{X}_1^{ee} \widehat{X}_3^{ee} \sin^2 \widetilde F_{31}^{}
- 4 \widehat{X}_2^{ee} \widehat{X}_3^{ee} \sin^2 \widetilde F_{32}^{}
\; ,
\end{eqnarray}
where $\widehat{X}_i^{ee} \equiv
\widetilde{X}_i^{ee}/\left(VV^{\dagger}\right)_{ee}^{}$ (for $i=1,2,3$),
and $\widetilde F_{ij} = 1267\times \widetilde \Delta_{ij}^{}L/E$
with $\widetilde \Delta_{ij}^{}$ being in unit of ${\rm eV}^2$, $L$
being in unit of km and $E$ being in unite of MeV (for $ij=21,31,32$).
It is easy to check that $\widehat{X}_1^{ee} + \widehat{X}_2^{ee} +
\widehat{X}_3^{ee} =1$ holds. In the absence of both UV and matter
effects, one is therefore left with $\widehat{X}_i^{ee} =
|U^{}_{ei}|^2$, depending only on $\theta^{}_{12}$ and $\theta^{}_{13}$.

The above equations tell us that once the eigenvalues $\widetilde{E}_i$
are figured out, it will be straightforward to obtain the explicit expression
of $\widetilde P (\overline\nu_{e}^{} \to \overline\nu^{}_{e})$.
Since the antineutrino beam energy $E$ is only around a few MeV,
one may calculate the eigenvalues of
$\widetilde{\cal H}$ by expanding them in terms of the small parameters
\begin{eqnarray}
\alpha \equiv \frac{\Delta_{21}^{}}{\Delta_{31}^{}} \; , \qquad
\beta \equiv \frac{2 \sqrt{2} \ G_{\rm F}^{} N_{e}^{}E}
{\Delta_{31}^{}} \; , \qquad
\gamma = \frac{\sqrt{2} \ G_{\rm F}^{} N_{n}^{}E}
{\Delta_{31}^{}} \;
\end{eqnarray}
with $\Delta_{ij}^{} \equiv m_i^{2} - m_j^2$ (for $ij =21, 31, 32$) in vacuum
and the small elements of $\kappa$. It is certainly a very good approximation
to take $N^{}_e \simeq N^{}_n$ in reality, so
$\beta \simeq 2 \gamma = A/\Delta_{31}$ with
$A \equiv 2 \sqrt{2} \ G_{\rm F} N_e^{} E$ being a common matter
parameter. Given
$A\sim 1.52 \times 10^{-4}~{\rm eV^2}~ Y_e^{}(\rho/{\rm g}/{\rm cm}^3)
(E/{\rm GeV}) \simeq 1.98 \times 10^{-4} {\rm eV}^2 (E/{\rm GeV})$
for $\rho \simeq 2.6 ~{\rm g/cm^3}$ and $E \sim 4$ MeV in
reactor antineutrino experiments, $\beta$ and $\gamma$ are actually much smaller
than $\alpha$ in magnitude:
\begin{eqnarray}
\alpha \hspace{-0.17cm} & \simeq & \hspace{-0.17cm}
3.1 \times 10^{-2} \times \frac{\Delta_{21}^{}}{7.5 \times
	10^{-5} {~\rm eV^2}} \times \frac{\pm 2.4 \times 10
	^{-3} {~\rm eV^2}}{\Delta_{31}^{}} \;, \nonumber\\
\beta \hspace{-0.17cm} & \simeq & \hspace{-0.17cm}
3.3 \times 10^{-4} \times \frac{E}{4 {~\rm MeV}}
\times \frac{\pm 2.4 \times 10
	^{-3} {~\rm eV^2}}{\Delta_{31}^{}} \;, \nonumber\\
\gamma \hspace{-0.17cm} & \simeq & \hspace{-0.17cm}
1.6 \times 10^{-4} \times \frac{E}{4 {~\rm MeV}}
\times \frac{\pm 2.4 \times 10
	^{-3} {~\rm eV^2}}{\Delta_{31}^{}} \; ,
\end{eqnarray}
in which the ``$\pm$" signs of $\Delta^{}_{31}$ stand for the
normal mass ordering (NMO) and inverted mass ordering (IMO) of
three neutrinos, respectively.
It is clear that $\beta\sim \gamma \sim {\cal {O}} (\alpha^2)$ holds.
As for the small UV parameters, we take $\kappa_{11}^{} \sim
\kappa_{22}^{} \sim \kappa_{33}^{} \sim \kappa_{21}^{} \sim
\kappa_{31}^{} \sim \kappa_{32}^{} \sim {\cal {O}} (\alpha)$
as a reasonable assumption \cite{Antusch2006}.
Now the effective Hamiltonian in Eq. (4) can be expressed as
\begin{eqnarray}
\widetilde {\cal H}
\hspace{-0.17cm} & = & \hspace{-0.17cm} E_1^{} {\bf 1} +
\frac{\Delta_{31}^{}}{2 E} U^{\rm T} \Omega U^{*}\; ,
\end{eqnarray}
where $\Omega$ is a dimensionless matrix containing both UV
and matter effects:
\begin{eqnarray}
\Omega = U^{*} \begin{pmatrix} 0 & 0 & 0 \cr 0 & \alpha & 0 \cr 0 & 0 & 1
\end{pmatrix} U^{\rm T} -\left({\bf 1} - \kappa\right)^{\rm T} \begin{pmatrix}
\beta - \gamma &0 & 0 \cr 0 & - \gamma & 0 \cr 0 & 0 & -\gamma
\end{pmatrix}\left({\bf 1} - \kappa\right)^{*} \;.
\end{eqnarray}
By making some analytical approximations, one may first calculate the
eigenvalues of $\Omega$ and then figure out the eigenvalues of
$\widetilde {\cal H}$. After a straightforward but tedious exercise,
we arrive at the expressions of the eigenvalues $\lambda_i^{}$ of
$\Omega$ in matter as follows:
\begin{eqnarray}
\lambda_1^{}
\hspace{-0.17cm} & \simeq & \hspace{-0.17cm}
-\beta |U_{e1}^{}|^2 + \gamma + \frac{1}{2} \left(\xi^{}_1 -\frac{
\xi_3^{} + 2 \beta^2 |U_{e1}^{}|^2 |U_{e2}^{}|^2}{\alpha}\right)
\; ,\nonumber \\
\lambda_2^{}
\hspace{-0.17cm} & \simeq & \hspace{-0.17cm}
\alpha -\beta |U_{e2}^{}|^2
+ \gamma + \frac{1}{2} \left(\xi^{}_1 +\frac{
\xi_3^{} + 2 \beta^2 |U_{e1}^{}|^2 |U_{e2}^{}|^2}{\alpha}\right)
\; ,\nonumber \\
\lambda_3^{}
\hspace{-0.17cm} & \simeq & \hspace{-0.17cm}
1 - \beta|U_{e3}^{}|^2  + \gamma - \xi^{}_2 \; ,
\end{eqnarray}
where $\xi^{}_i$ (for $i=1,2,3$) measure the effect of indirect UV:
\begin{eqnarray}
\xi^{}_1 \hspace{-0.17cm} &=& \hspace{-0.17cm}
\beta \kappa_{11}^{} (1 - |U_{e3}^{}|^2) -
\gamma \left[ \kappa_{11}^{} \left(1 - |U_{e3}^{}|^2\right) +
\kappa_{22}^{} \left(1 - |U_{\mu 3}^{}|^2\right) + \kappa_{33}^{}
\left(1 - |U_{\tau 3}^{}|^2\right)\right.
\nonumber\\
&& \hspace{-0.17cm}\left.
- 2{\rm Re} \left(\kappa_{21}^{}
U_{e3}^{} U_{\mu 3}^{*} + \kappa_{31}^{} U_{e3}^{}
U_{\tau 3}^{*} + \kappa_{32}^{}
U_{\mu 3}^{} U_{\tau 3}^{*}\right) \right] \;,
\nonumber\\
\xi^{}_2 \hspace{-0.17cm} &=& \hspace{-0.17cm}
-\beta \kappa_{11}^{} |U_{e3}^{}|^2 +
\gamma \left[ \kappa_{11}^{} |U_{e3}^{}|^2 +
\kappa_{22}^{} |U_{\mu 3}^{}|^2 + \kappa_{33}^{}
|U_{\tau 3}^{}|^2 \right.
\nonumber\\
&& \hspace{-0.17cm}\left.
+ 2{\rm Re} \left(\kappa_{21}^{}
U_{e3}^{} U_{\mu 3}^{*} + \kappa_{31}^{} U_{e3}^{}
U_{\tau 3}^{*} + \kappa_{32}^{}
U_{\mu 3}^{} U_{\tau 3}^{*}\right) \right] \;,
\nonumber\\
\xi^{}_3 \hspace{-0.17cm} &=& \hspace{-0.17cm}
\alpha \beta \kappa_{11}^{}
\left(|U_{e2}^{}|^2 - |U_{e1}^{}|^2\right) +
\alpha \gamma \left\{\kappa_{11}^{} \left(|U_{e1}^{}|^2
- |U_{e2}^{}|^2\right) \right.
\nonumber\\
&& \hspace{-0.17cm}
+ \kappa_{22}^{} \left(|U_{\mu 1}^{}|^2
- |U_{\mu 2}^{}|^2\right) +\kappa_{33}^{} \left(|U_{\tau 1}^{}|^2 -
|U_{\tau 2}^{}|^2\right)
\nonumber\\
&& \hspace{-0.17cm}
+ 2{\rm Re} \left(\kappa_{21}^{}
U_{e3}^{} U_{\mu 3}^{*} + \kappa_{31}^{} U_{e3}^{}
U_{\tau 3}^{*} + \kappa_{32}^{} U_{\mu 3}^{} U_{\tau 3}^{*}\right)
\nonumber\\
&& \hspace{-0.17cm}
-4 {\rm Re} \left[\kappa_{21}^{}\left(U_{e3}^{} U_{\tau 2}^{}
- U_{e2}^{} U_{\tau 3}^{}\right)\left(U_{\mu 3}^{*} U_{\tau 2}^{*}
- U_{\mu 2}^{*} U_{\tau 3}^{*}\right) \right.
\nonumber\\
&& \hspace{-0.17cm}
+ \kappa_{31}^{}\left(U_{e2}^{} U_{\mu 3}^{}
- U_{e3}^{} U_{\mu 2}^{}\right)\left(U_{\mu 3}^{*} U_{\tau 2}^{*}
- U_{\mu 2}^{*} U_{\tau 3}^{*}\right)
\nonumber\\
&& \hspace{-0.17cm} \left.\left.
+ \kappa_{32}^{}\left(U_{e3}^{} U_{\mu 2}^{}
- U_{e2}^{} U_{\mu 3}^{}\right)\left(U_{e 3}^{*} U_{\tau 2}^{*}
- U_{e 2}^{*} U_{\tau 3}^{*}\right) \right]\right\} \;.
\end{eqnarray}
One can see that in $\xi^{}_i$ the six UV parameters $\kappa^{}_{ij}$
are all entangled with the two matter parameters $\beta$ and $\gamma$,
implying that switching off the terrestrial matter effects will
automatically remove the indirect UV effect from $\lambda^{}_i$.
This important observation tells us that it will be much harder to
probe indirect UV for a low-energy oscillation experiment,
because the latter involves much smaller terrestrial matter effects.
Note that $\xi^{}_3$ is more suppressed in magnitude than $\xi^{}_1$ and
$\xi^{}_2$, but it cannot be ignored in
the expressions of $\lambda^{}_1$ and $\lambda^{}_2$ since
the combination $\xi^{}_3/\alpha$ should be comparable with the
$\xi^{}_1$ term in Eq. (13). With the help of Eq. (13), the eigenvalues
of $\widetilde {\cal H}$ can be directly obtained from
$\widetilde {E}_i^{} =E_1^{} + \lambda^{}_i \Delta_{31}^{}/\left(2E\right)$.
The three effective neutrino mass-squared
differences $\widetilde\Delta_{ij}^{}$ defined below Eq. (7) turn out to be
\begin{eqnarray}
\widetilde \Delta^{}_{21}
\hspace{-0.17cm}&\simeq&\hspace{-0.17cm}
\Delta_{31}^{} \left[\alpha +
\beta \left(|U_{e1}^{}|^2 - |U_{e2}^{}|^2 \right) + \frac{1}{\alpha}
\left(\xi_3^{} + 2 \beta^2 |U_{e1}^{}|^2 |U_{e2}^{}|^2\right)\right] \;,
\nonumber\\
\widetilde \Delta^{}_{31}
\hspace{-0.17cm}&\simeq&\hspace{-0.17cm}
\Delta_{31}^{} \left[1 + \beta \left(|U_{e1}^{}|^2 -
|U_{e3}^{}|^2 \right) - \frac{1}{2}
\left(\xi_1^{} + 2 \xi_2^{}\right) + \frac{1}{2 \alpha}
\left(\xi_3^{} + 2 \beta^2 |U_{e1}^{}|^2 |U_{e2}^{}|^2\right) \right] \;,
\nonumber\\
\widetilde \Delta^{}_{32}
\hspace{-0.17cm}&\simeq&\hspace{-0.17cm}
\Delta_{31}^{} \left[ 1 - \alpha + \beta \left(|U_{e2}^{}|^2 -
|U_{e3}^{}|^2 \right) -\frac{1}{2} \left(\xi_1^{} + 2 \xi_2^{}\right)
- \frac{1}{2 \alpha}
\left(\xi_3^{} + 2 \beta^2 |U_{e1}^{}|^2 |U_{e2}^{}|^2\right) \right] \;.
\end{eqnarray}
One can see that $\widetilde{\Delta}^{}_{21} = \widetilde{\Delta}^{}_{31} -
\widetilde{\Delta}^{}_{32}$ holds to the accuracy of the approximations
made above.

For simplicity, we are going to use $\widetilde {\cal H}^{\prime} =
\widetilde {\cal H} - E_1^{}{\bf 1}$ to calculate the probability of
$\overline\nu_e^{} \to \overline\nu_e^{}$ oscillations in the following,
since such a shift of $\widetilde {\cal H}$ does not affect any physics
under discussion. The results of $Y_i^{ee}$ and $N^{}_{ij}$
(for $i,j=1,2,3$) are listed in the Appendix. Then $\widetilde X_i^{ee}$
can be explicitly figured out with the help of Eq. (6). As a result,
the analytical approximations of
$\widehat X_i^{ee}$ defined below Eq. (8) turn out to be
\begin{eqnarray}
\widehat X_1^{ee} \hspace{-0.17cm} &\simeq&
\hspace{-0.17cm} |U_{e1}^{}|^2\left(1 + 2 \beta
|U_{e3}^{}|^2\right) + \frac{1}{2 \alpha}
\left(4 \beta |U_{e1}^{}|^2 |U_{e2}^{}|^2 - \xi^{}_4\right) -
\frac{\left(|U_{e1}^{}|^2 - |U_{e2}^{}|^2\right)}{\alpha^2}
\left(3 \beta^2 |U_{e1}^{}|^2 |U_{e2}^{}|^2 + \frac{\xi_3^{}}{2}\right)
\;,\nonumber\\
\widehat X_2^{ee} \hspace{-0.17cm} &\simeq&
\hspace{-0.17cm} |U_{e2}^{}|^2 \left(1 + 2 \beta
|U_{e3}^{}|^2\right)- \frac{1}{2 \alpha}
\left(4 \beta |U_{e1}^{}|^2 |U_{e2}^{}|^2 - \xi^{}_4\right) +
\frac{ \left(|U_{e1}^{}|^2 - |U_{e2}^{}|^2\right)}{\alpha^2}
\left(3 \beta^2 |U_{e1}^{}|^2 |U_{e2}^{}|^2 + \frac{\xi_3^{}}{2}\right)
\;, \nonumber\\
\widehat X_3^{ee} \hspace{-0.17cm} &\simeq&
\hspace{-0.17cm} |U_{e3}^{}|^2\left[1 - 2 \beta
\left(1-|U_{e3}^{}|^2\right)\right]  \;,
\end{eqnarray}
in which
\begin{eqnarray}
\xi^{}_4 \hspace{-0.17cm} &=& \hspace{-0.17cm}
2\kappa_{11}^{} \left(\beta
-\gamma\right)\left(1-2|U_{e3}^{}|^2\right) + 4 \gamma {\rm Re} \left(\kappa_{21} U_{e3}^{}
U_{\mu 3}^{*} + \kappa_{31}^{}  U_{e3}^{}  U_{\tau3}^{*}\right)
- \xi_1^{} +|U_{e3}^{}|^2 \left(\xi_1^{} -2 \xi_2^{}\right)
\nonumber \\
&=& \hspace{-0.17cm}
\beta \kappa_{11}^{} (1 - |U_{e 3}^{}|^2)^2 + \gamma
\left[- \kappa_{11}^{} \left(1 - |U_{e 3}^{}|^2\right)^2 +
\kappa_{22}^{} \left(|U_{\tau 3}^{}|^2 - |U_{e 3}^{}|^2
|U_{\mu 3}^{}|^2\right) +\kappa_{33}^{} \left(|U_{\mu 3}^{}|^2
\right.\right. \nonumber\\
&& \hspace{-0.17cm}
\left.\left. - |U_{e 3}^{}|^2|U_{\tau 3}^{}|^2\right)
+ 2\left(1 - |U_{e 3}^{}|^2\right){\rm Re} \left(\kappa_{21}^{}
U_{e3}^{} U_{\mu 3}^{*} +\kappa_{31}^{} U_{e3}^{}
U_{\tau 3}^{*}\right) \right.
\nonumber\\
&& \hspace{-0.17cm}
\left.- 2\left(1 + |U_{e 3}^{}|^2\right){\rm Re}
\left(\kappa_{32}^{}U_{\mu 3}^{} U_{\tau 3}^{*}\right) \right] \;.
\end{eqnarray}
The explicit expression of $\widetilde{P}(\overline{\nu}^{}_e \to
\overline{\nu}^{}_e)$ can therefore be obtained from Eq. (8) with
the help of Eq. (16). However, we prefer a different form of
$\widetilde{P}(\overline{\nu}^{}_e \to \overline{\nu}^{}_e)$
whose oscillation terms depend on $\widetilde{\Delta}^{}_{21}$ and
$\widetilde{\Delta}^{}_* \equiv \widetilde{\Delta}^{}_{31} +
\widetilde{\Delta}^{}_{32}$ \cite{Wang}, because $\widetilde{\Delta}^{}_*$
is sensitive to the neutrino mass ordering in a more transparent way.
According to Eqs. (2) and (15), we have
\begin{eqnarray}
\widetilde\Delta_{21}^{} \hspace{-0.17cm} &\simeq&
\hspace{-0.17cm} \Delta_{21}^{} + A \cos 2 \theta_{12}^{} \cos
^2 \theta_{13}^{} + A \left(
\frac{A}{2\Delta_{21}^{}} \sin^2 2\theta_{12}^{} \cos^4 \theta_{13}^{}
+ \xi_3^{\prime}\right)
\;,\nonumber\\
\widetilde\Delta_{*}^{} \hspace{-0.17cm} &\simeq&
\hspace{-0.17cm} \Delta_{*}^{} + A \left(1 - 3 \sin^2\theta_{13}^{}
- \xi^{\prime}_1 - 2 \xi^{\prime}_2\right) \;,
\end{eqnarray}
where $\Delta^{}_{21}$ and $\Delta_*^{} \equiv \Delta_{31}^{} + \Delta_{32}^{}$
are the counterparts of $\widetilde{\Delta}^{}_{21}$ and
$\widetilde \Delta_*^{}$ in vacuum, $\beta \simeq 2\gamma$ has been used, and
\begin{eqnarray}
\xi^{\prime}_1 \hspace{-0.17cm} &=& \hspace{-0.17cm}
\frac{1}{2} \left[\kappa_{11}^{} \left(1 - |U_{e3}^{}|^2\right) -
\kappa_{22}^{} \left(1 - |U_{\mu 3}^{}|^2\right) - \kappa_{33}^{} \left(1 -
|U_{\tau 3}^{}|^2\right)
\right.\nonumber\\ && \left.
+ 2{\rm Re} \left(\kappa_{21}^{}
U_{e3}^{} U_{\mu 3}^{*} + \kappa_{31}^{} U_{e3}^{}
U_{\tau 3}^{*} + \kappa_{32}^{}
U_{\mu 3}^{} U_{\tau 3}^{*}\right) \right] \;,
\nonumber\\
\xi^{\prime}_2 \hspace{-0.17cm} &=& \hspace{-0.17cm}
\frac{1}{2} \left[- \kappa_{11}^{} |U_{e3}^{}|^2 +
\kappa_{22}^{} |U_{\mu 3}^{}|^2 + \kappa_{33}^{}
|U_{\tau 3}^{}|^2
+ 2{\rm Re} \left(\kappa_{21}^{}
U_{e3}^{} U_{\mu 3}^{*} + \kappa_{31}^{} U_{e3}^{}
U_{\tau 3}^{*} + \kappa_{32}^{}
U_{\mu 3}^{} U_{\tau 3}^{*}\right) \right] \;,
\nonumber\\
\xi^{\prime}_3 \hspace{-0.17cm} &=& \hspace{-0.17cm}
\frac{1}{2}
\left\{\kappa_{11}^{} \left(|U_{e2}^{}|^2 - |U_{e1}^{}|^2\right)
-\kappa_{22}^{} \left(|U_{\mu 2}^{}|^2 - |U_{\mu 1}^{}|^2\right)
-\kappa_{33}^{} \left(|U_{\tau 2}^{}|^2 -|U_{\tau 1}^{}|^2\right)
 + 2{\rm Re} \left(\kappa_{21}^{} U_{e3}^{} U_{\mu 3}^{*}\right.\right.
 \nonumber\\
 &&\left.\left. + \kappa_{31}^{} U_{e3}^{}
U_{\tau 3}^{*} + \kappa_{32}^{} U_{\mu 3}^{} U_{\tau 3}^{*}\right)
-4 {\rm Re} \left[\kappa_{21}^{}\left(U_{e3}^{} U_{\tau 2}^{}
- U_{e2}^{} U_{\tau 3}^{}\right)\left(U_{\mu 3}^{*} U_{\tau 2}^{*}
- U_{\mu 2}^{*} U_{\tau 3}^{*}\right)
\right.\right.\nonumber\\
&&\left.\left. +
\kappa_{31}^{}\left(U_{e2}^{} U_{\mu 3}^{}
- U_{e3}^{} U_{\mu 2}^{}\right)\left(U_{\mu 3}^{*} U_{\tau 2}^{*}
- U_{\mu 2}^{*} U_{\tau 3}^{*}\right)+ \kappa_{32}^{}\left(U_{e3}^{} U_{\mu 2}^{}
- U_{e2}^{} U_{\mu 3}^{}\right)\left(U_{e 3}^{*} U_{\tau 2}^{*}
- U_{e 2}^{*} U_{\tau 3}^{*}\right) \right]\right\} \;,
\nonumber\\
\xi^{\prime}_4 \hspace{-0.17cm} &=& \hspace{-0.17cm}
\frac{1}{2}\left[\kappa_{11}^{} \left(1 - |U_{e 3}^{}|^2\right)^2 +
\kappa_{22}^{} \left(|U_{\tau 3}^{}|^2 - |U_{e 3}^{}|^2
|U_{\mu 3}^{}|^2\right) +\kappa_{33}^{} \left(|U_{\mu 3}^{}|^2
- |U_{e 3}^{}|^2|U_{\tau 3}^{}|^2\right)
\right.\nonumber\\
&& \hspace{-0.17cm}
\left.+ 2\left(1 - |U_{e 3}^{}|^2\right){\rm Re} \left(\kappa_{21}^{}
U_{e3}^{} U_{\mu 3}^{*} +\kappa_{31}^{} U_{e3}^{}
U_{\tau 3}^{*}\right) - 2\left(1 + |U_{e 3}^{}|^2\right){\rm Re}
\left(\kappa_{32}^{}U_{\mu 3}^{} U_{\tau 3}^{*}\right) \right] \;.
\end{eqnarray}
Different from $\xi^{}_i$ (for $i=1,2,3,4$), $\xi^\prime_i$ are purely
the UV parameters. Such a treatment will allow one to see the UV
effect in $\widetilde{P}(\overline{\nu}^{}_e \to \overline{\nu}^{}_e)$
more clearly. In Figure 1 we present a numerical illustration of
$\xi_i^{\prime}$ by inputting the $3\sigma$ ranges of the neutrino
oscillation parameters for the NMO case \cite{deSalas:2017kay}
and choosing the reasonable ranges of the UV parameters
(i.e., $\theta^{}_{ij} \lesssim 7.5^\circ$ and $\delta^{}_{ij} \in
[0, 2\pi)$ for $i=1,2,3$ and $j=4,5,6$). It is obvious that the
magnitudes of $\xi_i^{\prime}$ are either of the same order as
$\alpha$ or much smaller. Since the allowed ranges of $|\xi^\prime_i|$
in the IMO case are very similar to those in the NMO case,
they will not necessarily be shown here.
\begin{figure}[t]
	\centering
	\includegraphics[]{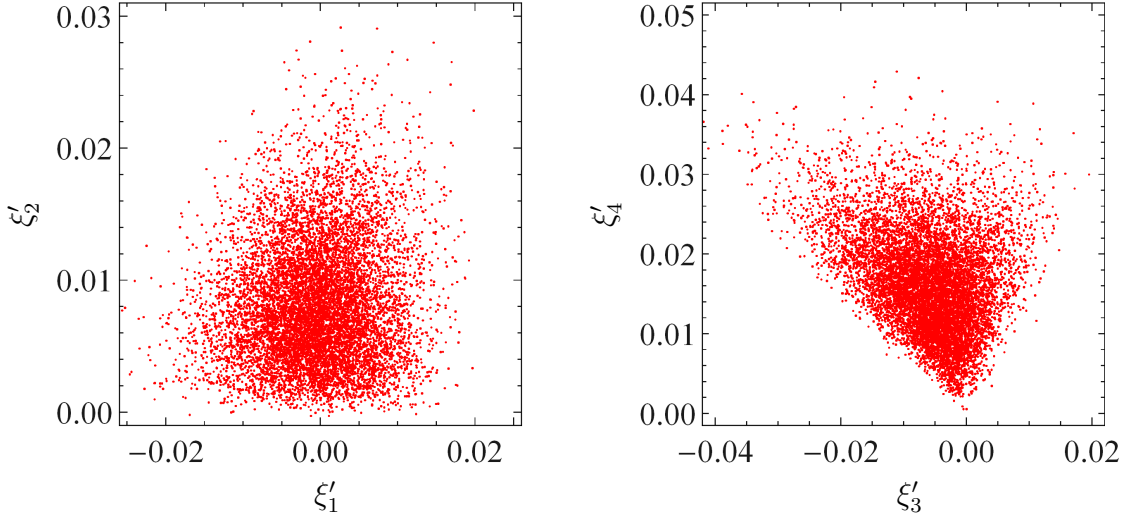}
	\caption{An illustation of $\xi_i^{\prime}$ given in Eq. (19) by
		inputting the 3$\sigma$ ranges of the six neutrino oscillation
		parameters (i.e., $\Delta_{21}^{}$, $\Delta_{31}^{}$, $\theta_{12}^{}$,
		$\theta_{13}^{}$, $\theta_{23}^{}$ and $\delta_{13}^{}$) for the
		NMO case \cite{deSalas:2017kay} and choosing the UV parameters
		in the ranges $\theta_{ij}^{} \lesssim 7.5^{\circ}$
		and $\delta_{ij}^{} \in [0, 2\pi)$ (for $i=1,2,3$ and $j=4,5,6$).}
\end{figure}

Now let us focus on the probability of $\overline \nu_e^{} \to \overline
\nu_e^{}$ oscillations. In vacuum we have the elegant expression
$ P (\overline\nu_e^{} \to
\overline\nu_e^{}) = 1 - P_0^{} - P_*^{}$ with \cite{Wang}
\begin{eqnarray}
{P}_0^{} \hspace{-0.17cm} &=& \hspace{-0.17cm}
\sin^2 2 \theta_{12}^{} \cos^4 \theta_{13}^{} \sin^2 F_{21}^{} \;,
\nonumber\\
{P}_*^{} \hspace{-0.17cm} &=& \hspace{-0.17cm} \frac{1}{2}
\sin^2 2 \theta_{13}^{} \left(1 - \cos F_*^{} \cos F_{21}^{}
+ \cos 2 \theta_{12}^{} \sin F_*^{} \sin F_{21}\right) \; ,
\end{eqnarray}
in which the term proportional to $\sin F^{}_*$ is sensitive to
the neutrino mass ordering. In matter with the UV effect,
the expression of $\widetilde P(\overline\nu_e^{} \to \overline\nu_e^{})$
shown in Eq. (8) can anagously be rewritten as
$\widetilde P (\overline\nu_e^{} \to \overline\nu_e^{})
= 1 - \widetilde P_0^{} - \widetilde P_*^{}$, where
$\widetilde P_0^{}$ represents the $\widetilde{\Delta}^{}_{21}$-
triggered oscillation and $\widetilde P_*^{}$ stands for the
$\widetilde{\Delta}^{}_{*}$-triggered oscillation.
Taking account of Eqs. (2), (8), (15), (16) and (18), we first define
\begin{eqnarray}
\widetilde{P}_0^{} \hspace{-0.17cm} &=&
\hspace{-0.17cm} P_0^{} + P_0^{\rm M_1^{}} + P_0^{\rm M_2^{}} +
P_0^{\rm UV} \;,
\nonumber\\
\widetilde{P}_*^{} \hspace{-0.17cm} &=&
\hspace{-0.17cm} P_*^{} + P_*^{\rm M_1^{}} + P_*^{\rm M_2^{}} +
P_*^{\rm UV} \;,
\end{eqnarray}
and then obtain
\begin{eqnarray}
{P}_0^{\rm M_1^{}} \hspace{-0.17cm} &\simeq& \hspace{-0.17cm}
A \sin^2 2 \theta_{12}^{} \cos 2\theta_{12}^{}
\cos^6 \theta_{13}^{} \left(1267 \frac{L}{E}
\sin 2 F_{21}^{} - \frac{2}{\Delta_{21}^{}}
\sin^2 F_{21}^{}\right) \;,
\nonumber\\
{P}_0^{\rm M_2^{}} \hspace{-0.17cm} &\simeq& \hspace{-0.17cm}
 A^2 \sin^2 2 \theta_{12}^{} \cos^8 \theta_{13}^{}
\left\{1267 \frac{L}{E} \left[\frac{1}{
2 \Delta_{21}^{}} \left(1 - 5 \cos^2 2 \theta_{12}^{}\right)
\sin 2 F_{21}^{} \right.\right.
\nonumber\\
&&\hspace{-0.17cm}\left.\left. + 1267 \frac{L}{E} \cos^2 2
\theta_{12}^{} \cos 2 F_{21}^{}\right] - \frac{1}{
\Delta_{21}^{2}} \left(1 - 4 \cos^2 2 \theta_{12}^{}\right)
\sin^2 F_{21}^{}\right\}
\nonumber\\
&& \hspace{-0.17cm}
+ \frac{A}{\Delta_{31}^{}} \sin^2 2 \theta_{12}^{}
\sin^2 2 \theta_{13}^{}
\cos^2 \theta_{13}^{} \sin^2 F_{21}^{}\;,
\nonumber\\
{P}_0^{\rm UV} \hspace{-0.17cm} &\simeq& \hspace{-0.17cm}
A \sin^2 2 \theta_{12}^{} \cos^2 \theta_{13}^{}\left[
1267 \xi_3^{\prime} \frac{L}{E} \cos^2 \theta_{13}^{}
\sin 2 F_{21}^{} +\frac{2}{\Delta_{21}^{}} \left( \xi_3^{\prime}
\cos 2 \theta_{12}^{} \cos^2 \theta_{13}^{}
\right.\right.
\nonumber\\
&& \left.\left. \hspace{-0.17cm}
+ \xi^{\prime}_4\right) \frac{\cos 2
\theta_{12}^{}}{ \sin^2 2 \theta_{12}^{}}\sin^2 F_{21}^{}\right] \; ;
\end{eqnarray}
and
\begin{eqnarray}
{P}_*^{\rm M_1^{}} \hspace{-0.17cm} &\simeq& \hspace{-0.17cm}
\frac{1}{2} A \sin^2 2 \theta_{13}^{} \left\{
1267 \frac{L}{E} \left[\left(1 + \cos^2 2 \theta_{12}^{}
\cos^2 \theta_{13}^{} -3 \sin^2 \theta_{13}^{}
\right) \sin F_{*}^{} \cos F_{21}^{} \right.\right.
\nonumber\\
&&\hspace{-0.17cm}\left.\left.
+ 2 \cos 2\theta_{12}^{}  \cos 2 \theta_{13}^{} \cos
F_{*}^{} \sin F_{21}^{}\right] + \frac{1}{\Delta_{21}^{}}
\sin^2 2 \theta_{12}^{} \cos^2 \theta_{13}^{}
\sin F_*^{} \sin F_{21}^{}\right\} \;,
\nonumber\\
{P}_*^{\rm M_2^{}} \hspace{-0.17cm} &\simeq& \hspace{-0.17cm}
\frac{1}{4} A^2 \sin^2 2 \theta_{13}^{} \left\{
1267 \frac{L}{E\Delta_{21}^{}}\sin^2 2
\theta_{12}^{}\cos^2\theta_{13}^{}\left[\left(3-7 \sin^2\theta_{13}^{}\right)
\cos F_*^{} \sin F_{21}^{} + 3\cos 2 \theta_{12}^{}
\right.\right.
\nonumber\\
&&\hspace{-0.17cm}\left.
\times\cos^2\theta_{13}^{}\sin F_*^{} \cos F_{21}^{}
\right] - \frac{3}{\Delta_{21}^{2}}
\sin^2 2 \theta_{12}^{} \cos 2 \theta_{12}^{}\cos^4 \theta_{13}^{}
\sin F_*^{} \sin F_{21}^{}  +\left(1267 \frac{L}{E}\right)^2 \left[
\right.
\nonumber\\
&&\hspace{-0.17cm}\left[1 + 3
\cos^2 2 \theta_{12}^{} - 2 \sin^2 \theta_{13}^{} \left(3 +
5 \cos^2 2\theta_{12}^{}\right) + \sin^4 \theta_{13}^{} \left(9 +
7\cos^2 2\theta_{12}^{}\right)\right]\cos F_*^{} \cos F_{21}^{}
\nonumber\\
&& \hspace{-0.17cm}
- \cos 2 \theta_{12}^{}\left[3 + \cos^2 2 \theta_{12}^{} - 2 \sin^2
\theta_{13}^{} \left(7 + \cos^2 2\theta_{12}^{}\right) +
\sin^4 \theta_{13}^{} \left(15 + \cos^2 2\theta_{12}^{}\right)\right]
\nonumber\\
&&\hspace{-0.17cm}
\left.\left. \times\sin F_*^{} \sin F_{21}^{}\right]
\right\}-\frac{A}{\Delta_{31}^{}}  \cos 2\theta_{13}^{}
\sin^2 2\theta_{13}^{} \left(1- \cos F^{}_* \cos F_{21}^{}
+\cos 2 \theta_{12}^{} \sin F_*^{}\sin F_{21}^{}\right)
\;,
\nonumber\\
{P}_*^{\rm{UV}} \hspace{-0.17cm} &\simeq& \hspace{-0.17cm}
\frac{1}{2} A \sin^2 2 \theta_{13}^{} \left\{
1267 \frac{L}{E} \left[\xi^{\prime}_3 \left(\cos F_*^{} \sin F_{21}^{}
 + \cos 2\theta_{12}^{} \sin F_*^{} \cos F_{21}^{}\right)
 - \left(\xi^{\prime}_1 + 2 \xi^{\prime}_2\right)
\right.\right.
\nonumber\\
&&\hspace{-0.17cm}
\left. \times \left(\cos 2 \theta_{12}^{}
\cos F_*^{} \sin F_{21}^{} + \sin F_*^{}
\cos F_{21}^{}\right)\right] - \frac{1}{\Delta_{21}^{}}
\left(\xi^{\prime}_3 \cos 2 \theta_{12}^{}
\right.
\nonumber\\
&&\hspace{-0.17cm}
\left.\left. + \frac{1}{\cos^2\theta_{13}^{}}\xi^{\prime}_4\right)
\sin F_*^{} \sin F_{21}^{}\right\} \;.
\end{eqnarray}
One can see that Eqs. (22) and (23) correspond to the matter-
and UV-induced corrections to the $P^{}_0$ and $P^{}_*$ terms,
respetively. Considering the smallness of $\sin^{}\theta_{13}^{}$,
let us simplify Eq. (23) to some extent as follows:
\begin{eqnarray}
{P}_*^{\rm M_1^{}} \hspace{-0.17cm} &=& \hspace{-0.17cm}
\frac{1}{2} A \sin^2 2 \theta_{13}^{} \left\{
1267 \frac{L}{E} \Bigr[\left(1 + \cos^2 2 \theta_{12}^{}
\cos^2 \theta_{13}^{} -3 \sin^2 \theta_{13}^{}
\right) \sin F_{*}^{} \cos F_{21}^{}
\right.\nonumber\\
&&\hspace{-0.17cm}\left.
+ 2 \cos 2\theta_{12}^{}  \cos 2 \theta_{13}^{} \cos
F_{*}^{} \sin F_{21}^{}\Bigr] + \frac{1}{\Delta_{21}^{}}
\sin^2 2 \theta_{12}^{} \cos^2 \theta_{13}^{}
\sin F_*^{} \sin F_{21}^{}\right\} \;,
\nonumber\\
{P}_*^{\rm M_2^{}} \hspace{-0.17cm} &=& \hspace{-0.17cm}
\frac{1}{4} A^2 \sin^2 2 \theta_{13}^{} \bigg\{
1267 \frac{3L}{E\Delta_{21}^{}}\sin^2 2
\theta_{12}^{}\left(
\cos F_*^{} \sin F_{21}^{} + \cos 2 \theta_{12}^{}
\sin F_*^{} \cos F_{21}^{}
\right) \nonumber\\
&&\hspace{-0.17cm}- \frac{3}{\Delta_{21}^{2}}
\sin^2 2 \theta_{12}^{} \cos 2 \theta_{12}^{}
\sin F_*^{} \sin F_{21}^{}  +\left(1267 \frac{L}{E}\right)^2 \Bigr[
\left(1 + 3\cos^2 2 \theta_{12}^{}\right)\cos F_*^{} \cos F_{21}^{}
\nonumber\\
&& \hspace{-0.17cm}
- \cos 2 \theta_{12}^{}(3 + \cos^2 2 \theta_{12}^{})
\sin F_*^{} \sin F_{21}^{}\Bigr]
\bigg\}-\frac{A}{\Delta_{31}^{}}
\sin^2 2\theta_{13}^{} \left(1 - \cos F^{}_* \cos F_{21}^{} \right.
\nonumber\\
&& \left. \hspace{-0.17cm}
 +\cos 2 \theta_{12}^{}
\sin F_*^{}\sin F_{21}^{}\right) \;,
\nonumber\\
{P}_*^{\rm{UV}} \hspace{-0.17cm} &=& \hspace{-0.17cm}
\frac{1}{2} A \sin^2 2 \theta_{13}^{} \left\{
1267 \frac{L}{E} \Bigr[\xi^{\prime}_3 \left(\cos F_*^{} \sin F_{21}^{}
+ \cos 2\theta_{12}^{} \sin F_*^{} \cos F_{21}^{}\right)
- \left(\xi^{\prime}_1 + 2 \xi^{\prime}_2\right)
\right.\nonumber\\
&&\hspace{-0.17cm}
\left. \times\left(\cos 2 \theta_{12}^{}
\cos F_*^{} \sin F_{21}^{} + \sin F_*^{}
\cos F_{21}^{}\right)\Bigr] - \frac{1}{\Delta_{21}^{}}
\left(\xi^{\prime}_3 \cos 2 \theta_{12}^{} + \xi^{\prime}_4\right)
\sin F_*^{} \sin F_{21}^{}\right\} \;. \hspace{0.5cm}
\end{eqnarray}
Note that the above analytical approximations are valid for
both the NMO and IMO cases, but can only be applied to the antineutrino oscillations.
As for the neutrino case, one ought to make the replacement of $\beta
\to -\beta$ and $\gamma\to -\gamma$. Some discussions are in order.
\begin{itemize}
\item     In the presence of indirect UV, our main analytical results
for $\widetilde P (\overline \nu_e^{} \to \overline \nu_e^{})$
are summarized in Eqs. (21), (22) and (24). We have done the expansions
up to ${\cal O} (\alpha^2)$ in our calculations, in which
$A/\Delta_{21}^{} \sim 1267 A L/E \sim 10^{-2} \sim {\cal O} (\alpha)$
is taken into account. The leading-order oscillation terms
$P_0^{{\rm M_1^{}}}$ and $P_*^{{\rm M_1^{}}}$ are consistent with those
obtained in Ref.~\cite{Wang}, where the UV effect was not considered.
In contrast, $P_0^{{\rm M}_2^{}}$, $P_*^{{\rm M_2^{}}}$,
$P_0^{{\rm UV}}$ and $P_{*}^{{\rm UV}}$ appear as the next-to-leading-order oscillation terms of $\widetilde P (\overline \nu_e^{} \to \overline \nu_e^{})$. Among these four new terms,
$P_0^{{\rm M_2^{}}}$ and $P_*^{{\rm M_2^{}}}$ describe the fine
terrestrial matter effects, and the other two characterize the
comparable or much smaller indirect UV effect.

\item     One can see that the UV effect is always smaller than
terrestrial matter effects, and their entanglement does not
appear until the next-to-leading-order oscillating terms are
taken into account. As for the UV-induced terms, $P_0^{\rm UV}$ is
modulated by the $\Delta^{}_{21}$-driven oscillation
while $P_*^{\rm UV}$ is the oscillation term related to
$\Delta_*^{}$ and might therefore affect the determination of
the neutrino mass ordering. Since both of them appear as the
next-to-leading-order terms as compared with $P_0^{{\rm M_1^{}}}$
and $P_*^{{\rm M_1^{}}}$, however, their effects must be strongly
suppressed.

{{
\item     In this paper, we only focus the indirect UV effect in which the masses of sterile
neutrinos are larger than the electroweak interaction scale. There is also another type of direct UV effect, where
sterile neutrinos can be produced and directly participate in the neutrino propagation process. Different from
the indirect UV effect considered here, sterile neutrinos in the direct UV framework will contribute additional
terms to the neutrino oscillation probability. In case that the oscillatory behavior can be observed, it will be tested or constrained
in the short baseline oscillations~\cite{Sterile,Gariazzo:2015rra,Gariazzo:2017fdh,Dentler:2017tkw,Fong:2017gke} for the mass-squared difference at around 1 eV$^2$
and at the JUNO-like experiment for the mass-squared difference from $10^{-5}$ eV$^2$ to $10^{-1}$ eV$^2$~\cite{JUNO}.
If these additional oscillations are averaged out, it will be similar to the indirect UV effect,
but with an additional constant term appeared as shown in Ref.~\cite{Fong2016,Li:2015oal}. According to Ref.~\cite{Parke:2015goa},
the limit on the corresponding active-sterile mixing will be relatively weaker in comparison to the indirect UV effect.
}}
\end{itemize}

\section{Numerical simulations}

In this section we shall first estimate the orders of magnitude of the oscillation terms associated with the UV and terrestrial matter effects using a JUNO-like detector, and then illustrate whether and how they can affect the neutrino mass ordering determination and precision measurements of $\Delta^{}_{21}$ and $\theta^{}_{12}$.
In our calculation the best-fit values of six active neutrino oscillation parameters are taken from a global analysis of current three-flavor oscillation experiments~\cite{deSalas:2017kay}, with $\Delta_{21} \simeq 7.56\times10^{-5}$ eV$^{2}$, $\sin^2\theta_{12}\simeq0.321$, $\Delta_{*} \simeq 5.024\times10^{-3}$ eV$^{2}$,
$\sin^2\theta_{13}\simeq0.022$, $\sin^2\theta_{23}\simeq0.430$ and $\delta\simeq252^{\circ}$ for the NMO case, and
with $\Delta_{21} \simeq 7.56\times10^{-5}$ eV$^{2}$, $\sin^2\theta_{12}\simeq0.321$, $\Delta_{*} \simeq -5.056\times10^{-3}$ eV$^{2}$,
$\sin^2\theta_{13}\simeq0.021$, $\sin^2\theta_{23}\simeq0.596$ and $\delta\simeq259^{\circ}$ for the IMO case.
The averaged terrestrial matter density along the reactor antineutrino trajectory is taken as $\rho \simeq 2.6 ~{\rm g/cm^3}$~\cite{Mocioiu:2000st}.
To illustrate the UV effect, we typically take $\theta_{14}^{}=\theta_{24}^{}=\theta_{34}^{}=
\theta_{15}^{}=\theta_{25}^{}=\theta_{35}^{}=
\theta_{16}^{}=\theta_{26}^{}=\theta_{36}^{}=5^{\circ}$,
$\delta_{14}^{}=\delta_{15}^{}=\delta_{16}^{}=120^{\circ}$,
$\delta_{24}^{}=\delta_{25}^{}=\delta_{26}^{}=60^{\circ}$ and
$\delta_{34}^{}=\delta_{35}^{}=\delta_{36}^{}=0^{\circ}$.
In addition, for the sensitivity calculation, we assume a JUNO-like 20-kiloton liquid scintillator detector with the energy resolution of 3\%/$\sqrt{E\,(\rm MeV)}$. The reactor power and baseline distributions are taken from Tab.~1 of Ref.~\cite{Li:2013zyd}, a total thermal power of 36 ${\rm GW}_{\rm th}$ and a weighted
baseline of 52.5 km. We assume the nominal running time of six years and 300 effective days per year in our numerical simulations. All the statistical and systematical setups are the same as those in Ref.~\cite{Wang}, where one can find all the simulation details. The only exception is that here we have enlarged the flux normalization uncertainty to 10\% in order to accommodate the reactor antineutrino anomaly and UV-induced zero-distance effect.

In Figure 2 we illustrate the numerical orders of magnitude of the matter-induced and UV-induced corrections to the oscillation probability, where the first and second rows are for the absolute and relative differences of the matter-induced correction respectively, and the third and fourth rows are for the absolute and relative differences of the UV-induced correction respectively. In the left and right panels we show the NMO and IMO cases respectively.
For illustration, we define the absolute error induced
by the UV and matter effects as
\begin{eqnarray}
\Delta {P}_{\rm UV}^{} \hspace{-0.17cm} &=& \hspace{-0.17cm}
\widetilde P (\overline\nu_e^{} \to \overline\nu_e^{}) -
\widetilde P (\overline\nu_e^{} \to \overline\nu_e^{},~ \kappa={\bf 0})
\nonumber\\
&\simeq&\hspace{-0.17cm} - \left(P_0^{\rm UV} + P_*^{\rm UV}\right) \;,
\nonumber\\
\Delta {P}_{\rm M}^{} \hspace{-0.17cm} &=& \hspace{-0.17cm}
\widetilde P (\overline\nu_e^{} \to \overline\nu_e^{}) -
\widetilde P (\overline\nu_e^{} \to \overline\nu_e^{},~ A=0)
\nonumber\\
&\simeq&\hspace{-0.17cm} - \left(P_0^{\rm M_1^{}} + P_0^{\rm M_2^{}}
+ P_0^{\rm UV} +P_*^{\rm M_1^{}} + P_*^{\rm M_2^{}} +
P_*^{\rm UV}\right)  \;,
\end{eqnarray}
where $\widetilde P (\overline\nu_e^{} \to \overline\nu_e^{},~
\kappa={\bf 0})$ denotes $\widetilde P (\overline\nu_e^{} \to
\overline\nu_e^{})$ in Eq.~(8) by taking $\kappa={\bf 0}$ with
${\bf 0}$ meaning all the elements of
$\kappa$ are zero (i.e., turning off the UV effect), and $\widetilde P
(\overline\nu_e^{} \to \overline\nu_e^{},~ A=0)$ stands for $\widetilde
P (\overline\nu_e^{} \to \overline\nu_e^{})$ with $A=0$
(i.e., back to the case in vacuum).
Compared to the left panel of Fig.~1 in Ref.~\cite{Wang}, here
the absolute difference $\Delta {P}_{\rm M}^{}$ is defined in a generic
framework with three active neutrinos and three heavy sterile neutrinos,
and it includes the interference terms of the UV and matter
potential parameters. The solid and dashed lines are
shown for the exact numerical calculation and analytical approximations
in Eqs. (21), (22) and (24), respectively.
From the first and second rows, we can observe that the absolute and
relative orders of magnitude of the matter-induced corrections can
reach the levels of 0.6\% and 4\% respectively, consistent
with those in Ref.~\cite{Wang} without the UV effect. On the other
hand, the absolute and relative orders of magnitude of the UV-induced
corrections are at most 0.02\% and 0.1\% according to the third and
fourth rows. This is because the UV effect is always entangled with
matter effects and appears at the next-to-leading order.
The same
conclusion can be drawn in Figure 3 where the individual terms of
the expansion done in Eqs. (21), (22) and (24) are illustrated. The
upper panels are for the leading oscillation terms $P_0^{{\rm
M_1^{}}}$ and $P_*^{{\rm M_1^{}}}$, and the four
next-to-leading terms are illustrated
in the lower panels. The left and right panels are shown for the NMO
and IMO cases respectively. To show how the UV-induced corrections
depend on the standard oscillation and UV parameters, we illustrate the
scattering plots of the UV-induced corrections in Figure 4 by varying
the six oscillation parameters
($\Delta_{21}^{},~\Delta_{31}^{},~\theta_{12}^{},~\theta_{13}^{},~\theta
_{23}^{},~\delta_{13}^{}$) within their 3$\sigma$ ranges for the NMO
case, and the UV parameters $\theta_{ij}^{}$ $(i=1,~2,~3,~j=4,~5,~6)$
and $\delta_{ij}^{}$ $(i=1,~2,~3,~j=4,~5,~6)$ in the ranges of
$[0,\;7.5^{\circ}]$ and $[0,\;360^{\circ}]$ respectively. The left and
right panels are shown for the exact numerical calculation and analytical
approximations respectively. We conclude that the
absolute magnitudes of the UV-induced corrections are within the
region of smaller than $0.05\%$.
\begin{figure}
	\centering
	\includegraphics[scale=0.9]{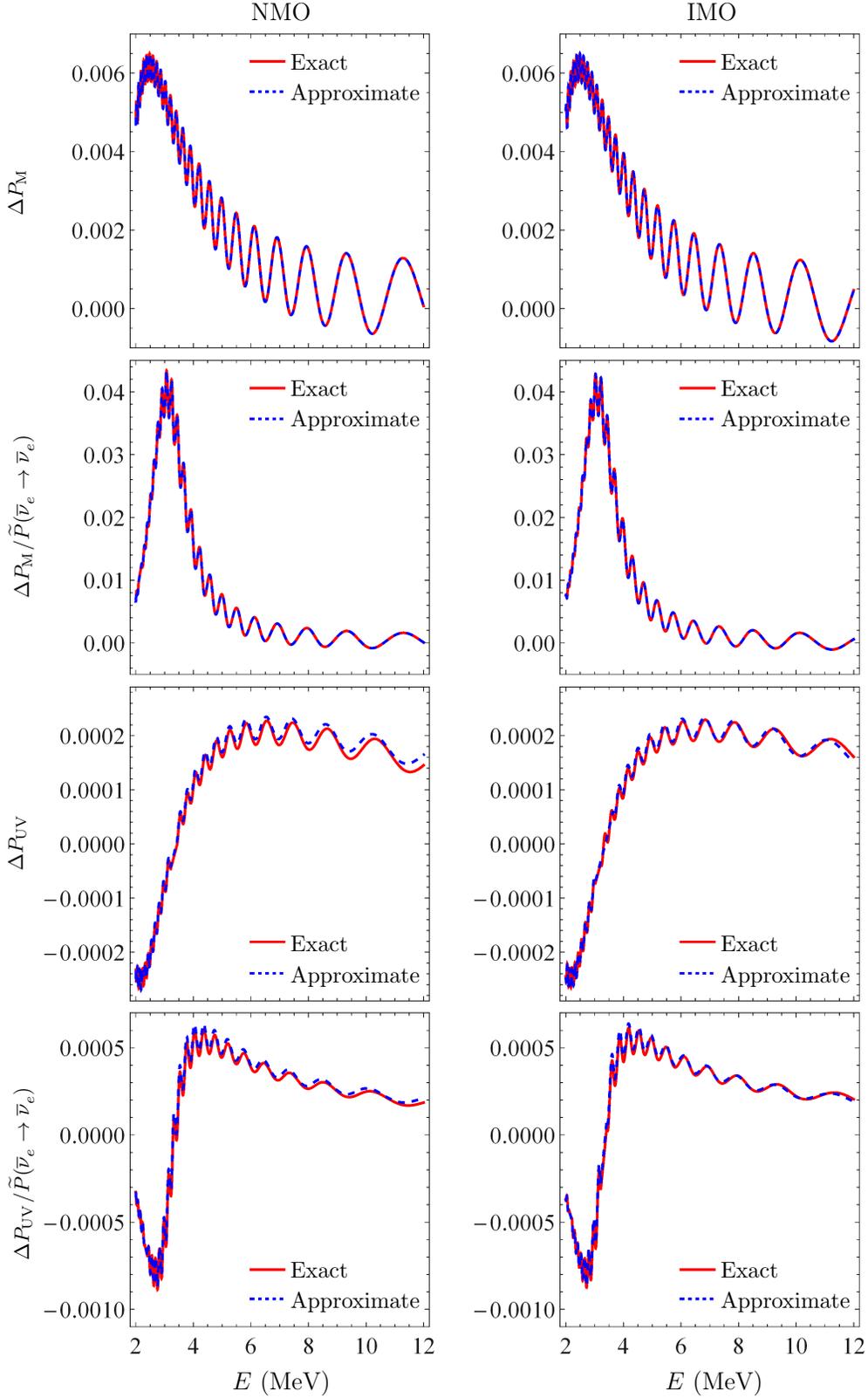}
	\caption{Numerical orders of magnitude of the matter-induced and UV-induced corrections to the oscillation probability, where the first (third) and second (fourth) rows are for the absolute and relative differences of the matter-induced (UV-induced) correction respectively. The left and right panels are shown for the NMO and IMO cases respectively.
The solid and dashed lines are shown for the exact numerical calculations and analytical approximations respectively.}
\end{figure}
\begin{figure}
	\centering
	\includegraphics[scale=0.9]{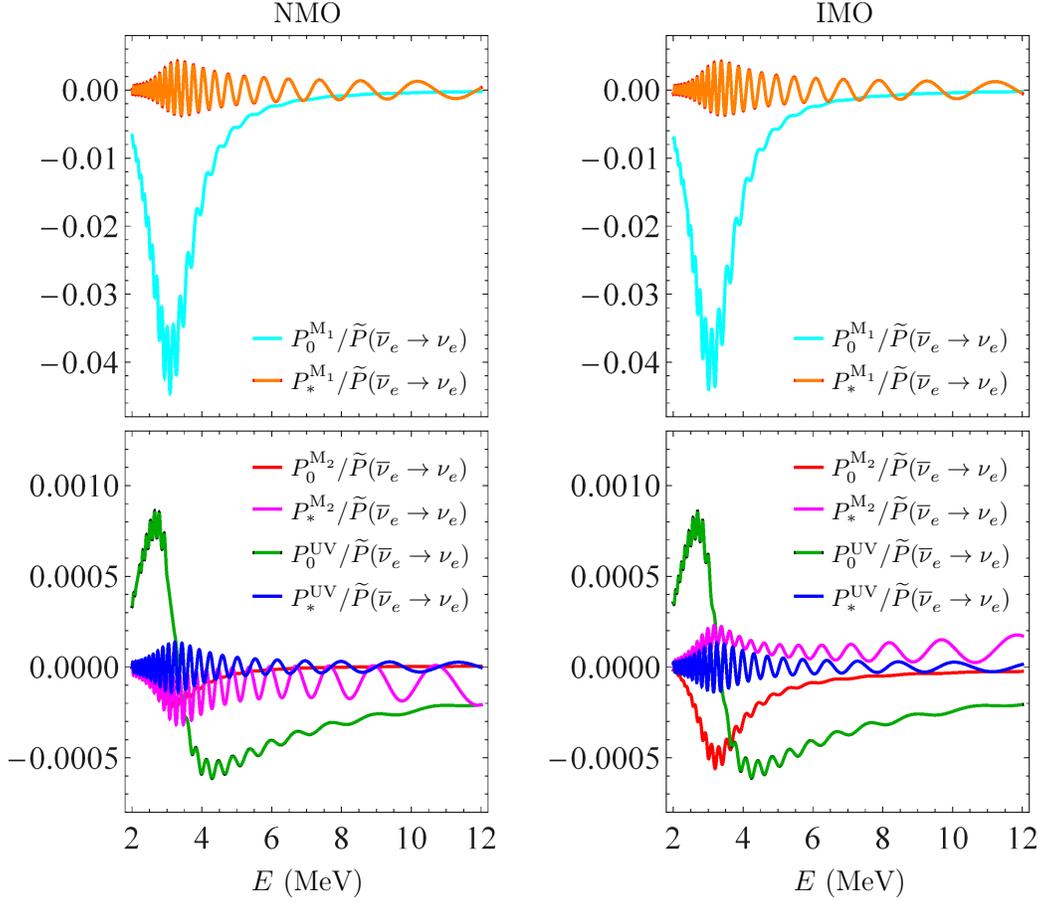}
	\caption{{{The relative}} numerical orders of magnitude of the individual expansion terms in Eqs. (22) and (24)
	{{to the analytical approximations of
$\widetilde P(\overline \nu_e^{} \to \overline\nu_e^{})$}}. The upper panels are for the leading oscillation terms $P_0^{{\rm M_1^{}}}$ and $P_*^{{\rm M_1^{}}}$, and the four next-to-leading terms are illustrated in the lower panels. The left and right panels are shown for the NMO and IMO cases respectively.
}
\end{figure}
\begin{figure}
	\centering
	\includegraphics[scale=0.95]{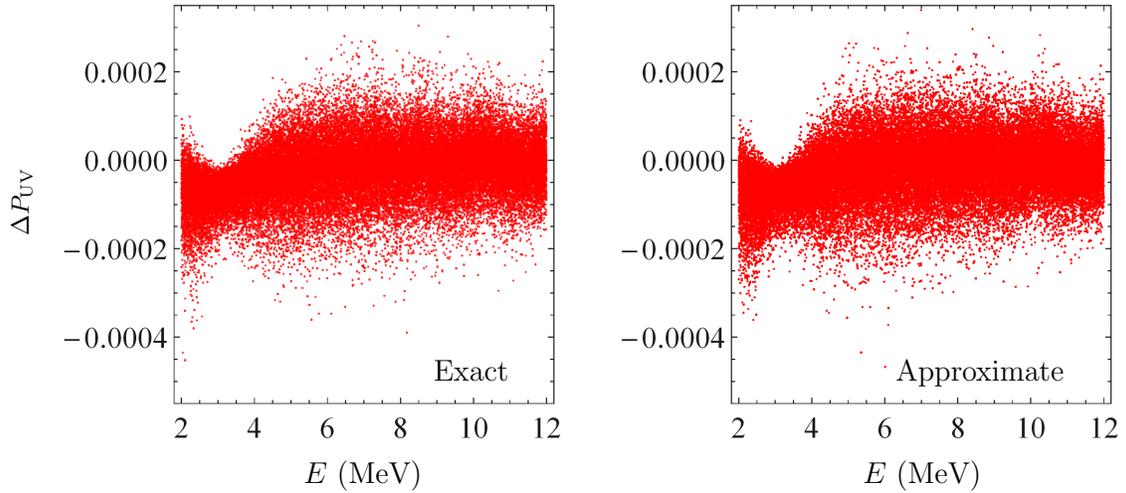}
	\caption{Scattering plots of the UV-induced corrections by varying the six oscillation parameters ($\Delta_{21}^{},~\Delta_{31}^{},~\theta_{12}^{},~\theta_{13}^{},~\theta_{23}^{},~\delta_{13}^{}$) within their 3$\sigma$ ranges for the NMO case, and the UV parameters $\theta_{ij}^{}$ $(i=1,~2,~3,~j=4,~5,~6)$ and $\delta_{ij}^{}$ $(i=1,~2,~3,~j=4,~5,~6)$ in the ranges of $[0,\;7.5^{\circ}]$ and $[0,\;360^{\circ}]$ respectively.
The left and right panels are shown for the exact numerical calculations and analytical approximations respectively.
	}
\end{figure}

In Figure 5 we illustrate the terrestrial matter (left panel) and UV (right panel) effects on the neutrino mass ordering sensitivity in the generic framework
of three active neutrinos and three heavy sterile neutrinos. In each panel the vertical distances of the black and red lines are defined as the sensitivity of the mass ordering 
{ {(i.e., $\Delta\chi^2_{\rm}=|\chi^2_{\rm min}(\rm NMO) -\chi^2_{\rm min}(\rm IMO)|$, where the least squares function $\chi^2$ is defined as in Eq.~(20) of Ref.~\cite{Wang} and 
$\chi^2_{\rm min}$ is the minimum of $\chi^2$ after the marginalization of all the oscillation and pull parameters)}}. 
The solid lines are for the case considering both the matter and UV effects and the dashed lines are the scenario of neglecting the matter effects (left panel) 
or neglecting the UV effect (right panel). Note that the red dashed line in the right panel has been horizontally shifted by $-0.35\times10^{-5}\;{\rm eV}^2$ 
to avoid the overlap of the curves. In the left panel, the inclusion of terrestrial matter effects can reduce $\Delta\chi^2_{\rm }$ by 0.61 from 9.89 to 9.28. 
This conclusion is consistent with that in Ref.~\cite{Wang} for the three neutrino mixing case ($\Delta\chi^2_{\rm }$ reduced by 0.64 from 10.28 to 9.64). 
The absolute value of $\Delta\chi^2_{\rm }$  is reduced mainly because the true three neutrino oscillation parameters have been changed to those in Ref.~\cite{deSalas:2017kay}. 
The size of $\Delta\chi^2_{\rm }$ reduction by 0.61 is non-negligible because it can be comparable with other systematic uncertainties. 
On the other hand, one can observe from the right panel that the inclusion of the UV effect only change $\Delta\chi^2_{\rm }$ from 9.31 to 9.28, 
resulting in a reduction of $\Delta\chi^2_{\rm }\simeq0.03$, which is much smaller than that of terrestrial matter effects. By randomly sampling the UV parameters $\theta_{ij}^{}$ $(i=1,~2,~3,~j=4,~5,~6)$ and $\delta_{ij}^{}$ $(i=1,~2,~3,~j=4,~5,~6)$ in the ranges of $[0,\;7.5^{\circ}]$ and $[0,\;360^{\circ}]$ respectively, we find that the variation of $\Delta\chi^2_{\rm }$ is well within the $\pm0.04$ range around 9.28, which demonstrates the robustness of the mass ordering measurement against the possible UV effect in the JUNO experiment.

Next we are going to discuss the UV and terrestrial matter effects in the precision measurement of $\theta^{}_{12}$ and $\Delta^{}_{21}$. In Figure 6 we illustrate the fitting results of $\theta^{}_{12}$ and $\Delta^{}_{21}$ where both the matter and UV effects are included in the measured neutrino spectrum but the matter effects (left panel) or the UV corrections (right panel) are neglected in the predicted neutrino spectrum. The red stars and blue circles are the true values and best-fit values of $\theta^{}_{12}$ and $\Delta^{}_{21}$, respectively.
From the left panel for the case of neglecting matter effects, one can observe that the best-fit values of $\theta^{}_{12}$ and $\Delta^{}_{21}$ deviate around 2.0$\sigma$ and 0.7$\sigma$ from their true values, with the parameter precisions of 0.63\% and 0.29\% respectively. The levels of deviations for the fitted $\theta^{}_{12}$ and $\Delta^{}_{21}$ are similar to those obtained in Ref.~\cite{Wang} where the three-flavor oscillation framework is considered. Thus terrestrial matter effects are of importance for future precision spectral measurements of reactor antineutrino oscillations.
Regarding the case of neglecting the UV effect as shown in the right panel, the deviations of the best-fit values for $\theta^{}_{12}$ and $\Delta^{}_{21}$ are within the size of $0.1~\sigma$ with the parameter precisions of 0.60\% and 0.27\% respectively. The parameter accuracies in the left panel are a little bit worse because additional marginalization has been performed for the UV parameters in the same regions as in Figure 4. Therefore the precision measurement of $\theta^{}_{12}$ and $\Delta^{}_{21}$ in the generic framework of three active neutrinos and three heavy sterile neutrinos turns out to be rather robust for the reasonable UV parameter space.

{{
Before finishing this section, we want to remark on the indirect UV effect in the accelerator neutrino experiments.
Different from the oscillation channel $\overline \nu_e^{} \to \overline \nu_e^{}$ for the reactor antineutrino experiments discussed here,
the indirect UV effect in long baseline accelerator neutrino experiments may be significant because the terrestrial matter effect becomes larger and its entanglement
with the indirect UV effect will also be non-negligible.
The additional mixing angles and CP-violating phases will induce multiple parameter degeneracy problem and the sensitivities to the neutrino mass ordering,
leptonic CP violation and the $\theta_{23}$ octant will be largely affected~\cite{Antusch:2009pm,Escrihuela:2015wra,Bekman:2002zk,
Ge:2016xya, Miranda:2016wdr,Escrihuela:2016ube,C:2017scx,Tang}. Taking the DUNE experiment as an example, the discovery potential for maximal CP violation
would be degraded from $6\sigma$ to the $3.7\sigma$ for seven years of nominal running if the indirect UV effect is considered~\cite{Tang}.
A robust method to remove the parameter degeneracy and have better sensitivities to the three neutrino oscillation and new physics effects
would be the combinations of accelerator neutrino experiments with different baselines, different neutrino energies and different neutrino oscillation channels~\cite{Tang}.}}
\begin{figure}
\begin{center}
\begin{tabular}{cc}
\includegraphics[bb=50 30 720 520, width=0.45\textwidth]{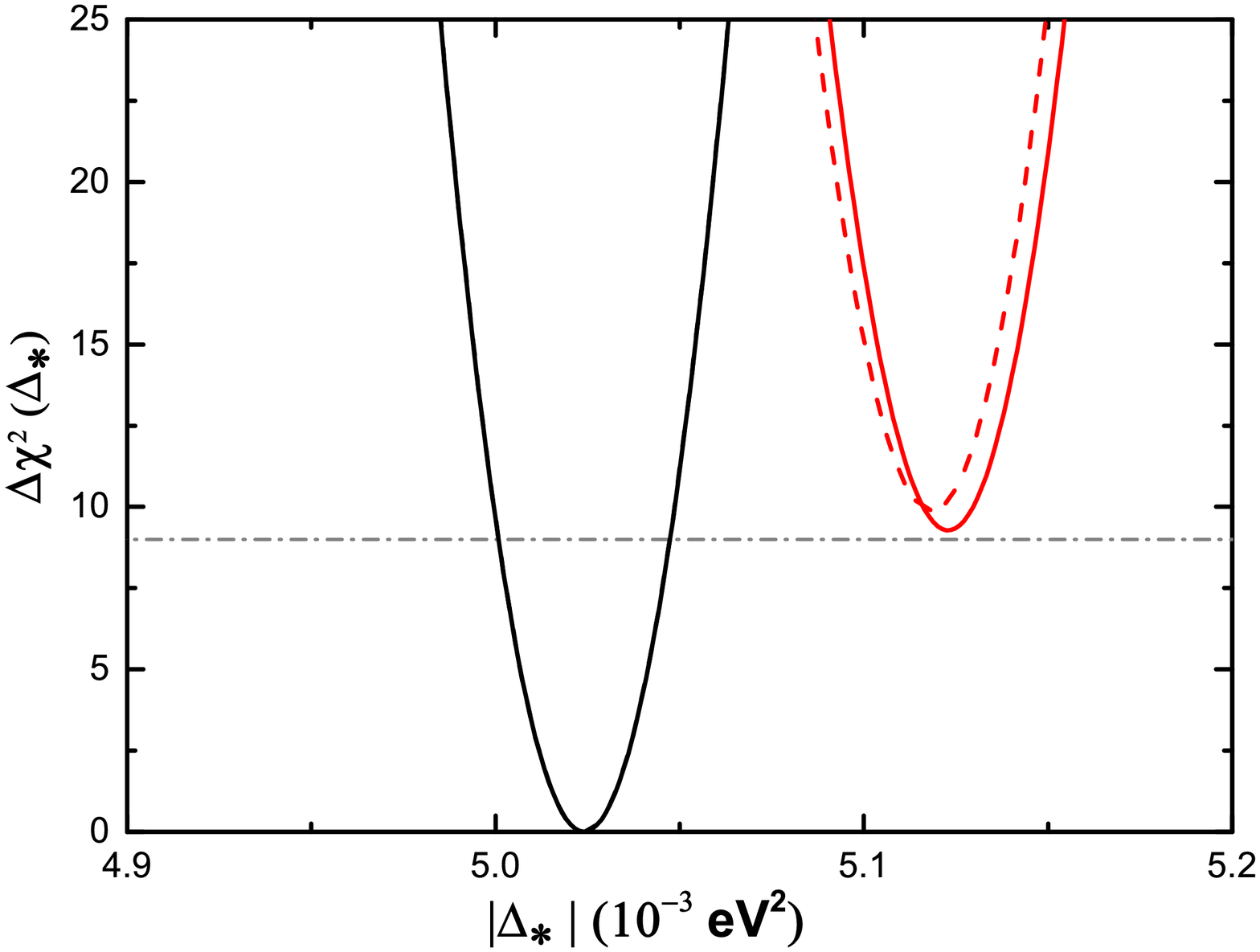}
&
\includegraphics[bb=50 30 720 520, width=0.45\textwidth]{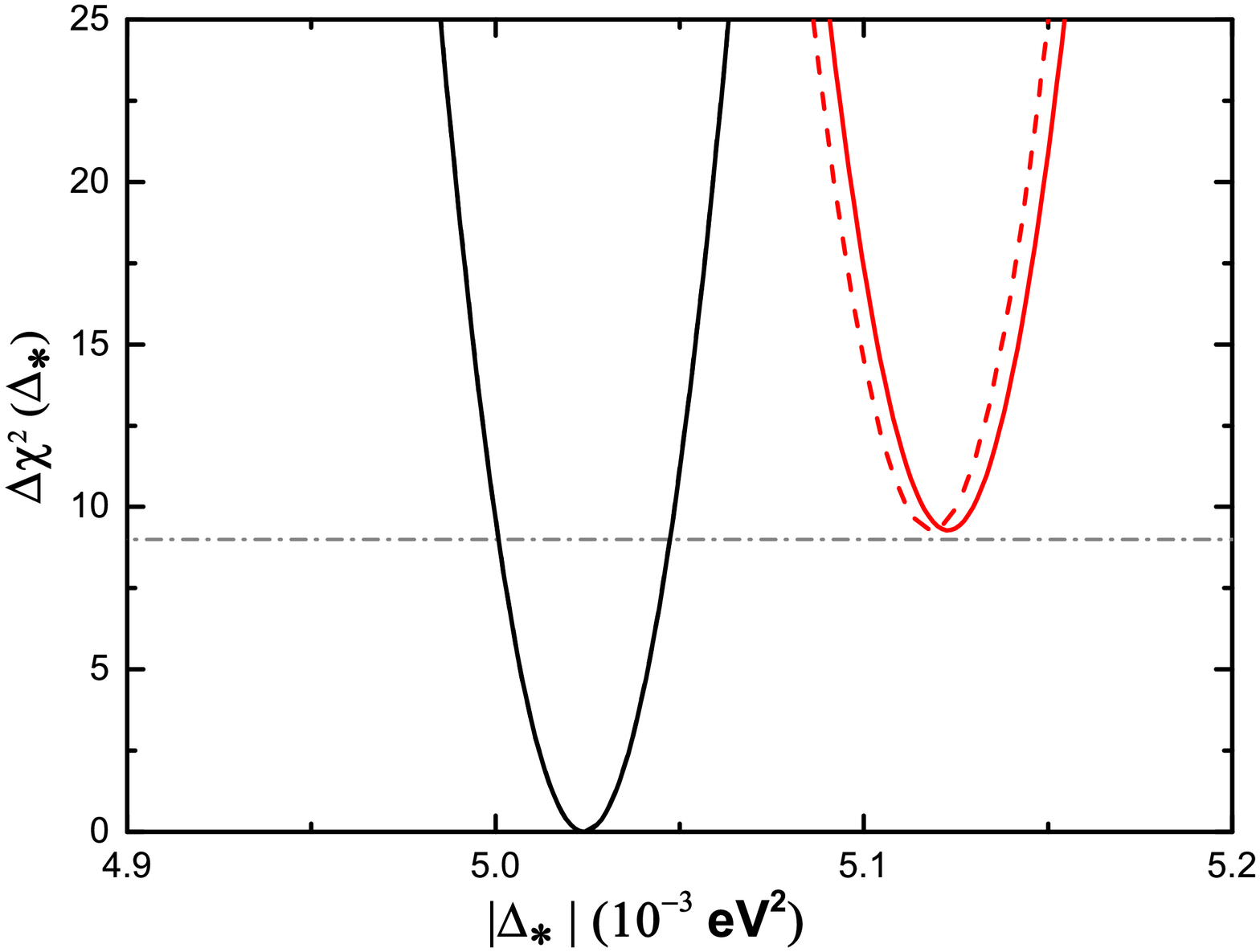}
\end{tabular}
\end{center}
\caption{Terrestrial matter (left panel) and UV (right panel) effects
on the neutrino mass ordering sensitivity in the generic framework
of three active neutrinos and three heavy sterile neutrinos.
The solid lines are for the case considering both matter and UV
effects{{, where the black and red ones come from the
fitting assuming the NMO and IMO cases of three active neutrinos,
respectively.}}
The dashed lines are the scenario of neglecting the matter effects (left panel) or neglecting the UV effect (right panel).
{{In each panel the vertical distances between the minima of the black and red lines are defined as
the sensitivity of the mass ordering (i.e., $\Delta\chi^2_{\rm}$)}.}
\label{fig:5}}
\end{figure}
\begin{figure}
\begin{center}
\begin{tabular}{cc}
\includegraphics[bb=50 30 720 520, width=0.45\textwidth]{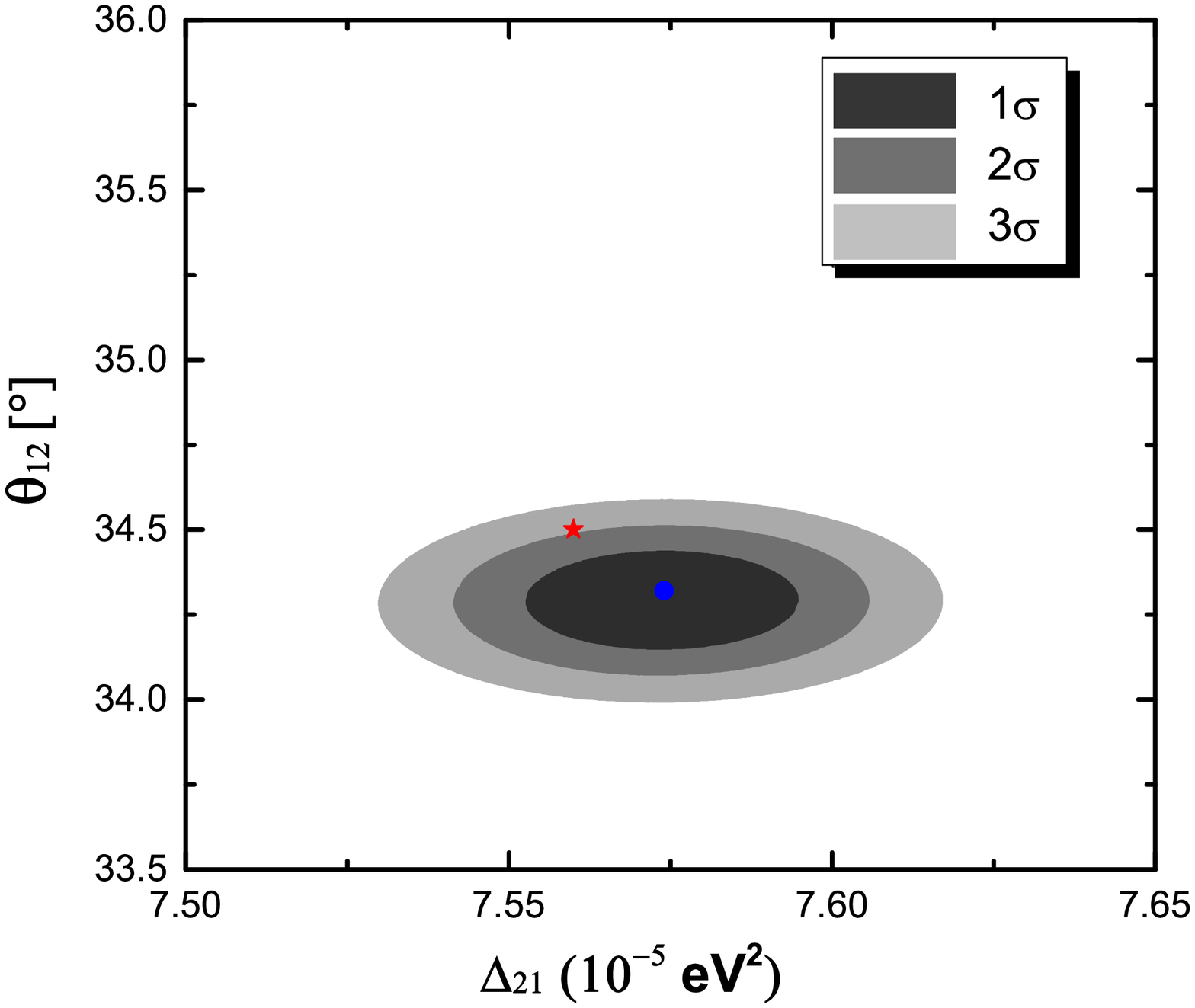}
&
\includegraphics[bb=50 30 720 520, width=0.45\textwidth]{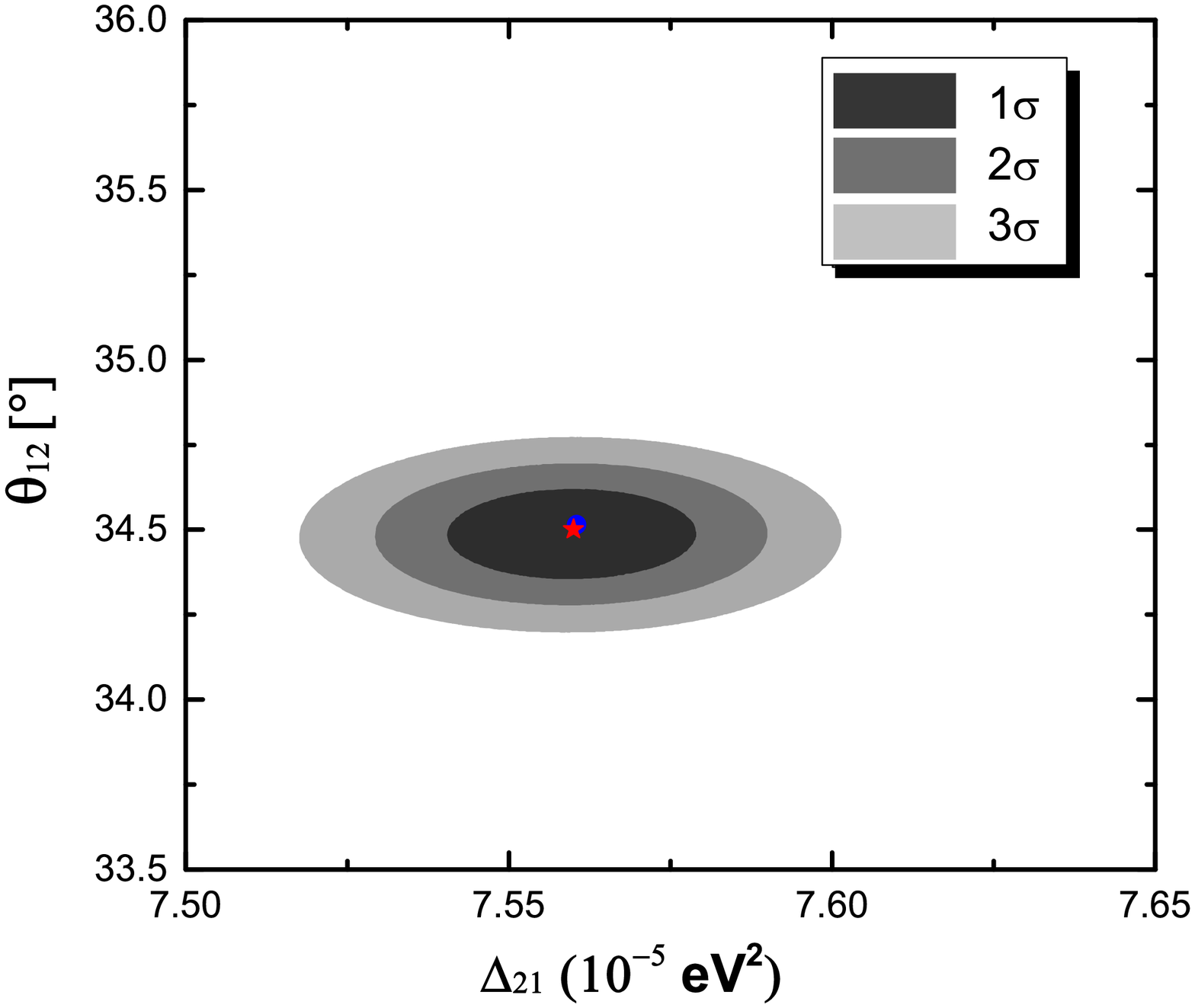}
\end{tabular}
\end{center}
\caption{Allowed regions of $\theta^{}_{12}$ and $\Delta^{}_{21}$ by neglecting the terrestrial matter (left panel) or UV (right panel) effects
in the predictions. Both effects are included in the measurements. The red stars and blue circles are the true values and best-fit values
of $\theta^{}_{12}$ and $\Delta^{}_{21}$, respectively.
\label{fig:6}}
\end{figure}

\section{Summary}

We have examined whether the JUNO-like reactor antineutrino oscillation
experiment can be used to probe the indirect UV effect caused by small
corrections of heavy sterile neutrinos to the $3\times 3$ PMNS matrix.
In this regard we have paid particular attention to how such an effect
is entangled with terrestrial matter effects in
$\overline{\nu}^{}_e \to \overline{\nu}^{}_e$ oscillations.
After deriving the oscillation probability in a good analytical approximation
for the antineutrino beam energy of a few MeV, we have
done some numerical simulations based on
the setup of a 20-kiloton JUNO-like liquid scintillator detector.
We find that the indirect UV effect is always smaller than terrestrial
matter effects, and their entanglement does not
appear until the next-to-leading-order oscillating terms
are taken into account. Two immediate conclusions turn out to be:
(a) indirect UV makes no difference in the JUNO-like experiment; and
(b) such an experiment's sensitivities to the neutrino
mass ordering and a precision measurement of $\theta^{}_{12}$ and
$\Delta^{}_{21}$ are robust.

Although the indirect UV effect is too small to be accessible in the
JUNO-like reactor-based antineutrino oscillation experiment, it may
be probed or constrained in some accelerator-based long-baseline neutrino
oscillation experiments. In either case terrestrial matter effects
should be carefully studied, so as to make them distinguishable
from the fundamental new physics effects.

\vspace{0.2cm}
\begin{flushleft}
{\large\bf Acknowledgments}
\end{flushleft}
One of us (ZZX) would like to thank H. P$\rm\ddot{a}$s for his warm hospitality
and useful discussions at the Technische Universit$\rm\ddot{a}$t Dortmund, where
this paper was finalized.
This work was supported in part by the National Natural Science Foundation of China under Grant No.~11305193 (YFL) and Grant No. 11775231 (ZZX and JYZ).

\vspace{0.3cm}
\begin{flushleft}
{\Large\bf Appendix}
\end{flushleft}

One may calculate $Y_k^{ee}=(V^{*} \widetilde {\cal H}^{k-1}V^{\rm T})^{ee}$
and $N^{}_{jk}$ (for $j, k = 1,2,3$) in Eqs. (6) and (7) with the replacements
$\widetilde {\cal H} \to \widetilde {\cal H}^{\prime} = \widetilde {\cal H} - E_1^{}{\bf 1}$,
and $\widetilde{E}^{}_i \to \widetilde{E}^{\prime}_i = \widetilde{E}^{}_i -
E^{}_1 = \lambda^{}_i \Delta_{31}^{}/\left(2E\right)$.
The explicit expressions of $Y_i^{ee}$ in our approximations are
\begin{eqnarray}
Y^{ee}_1 \hspace{-0.17cm} &\simeq& \hspace{-0.17cm}
\left(1 - \frac{\kappa_{11}^{}}{2}\right)^2 \;,
\nonumber\\
Y^{ee}_2 \hspace{-0.17cm} &\simeq& \hspace{-0.17cm}
\frac{\Delta_{31}^{}}{2 E} \left(1 - \frac{\kappa_{11}^{}}{2}\right)^2
\left[ |U_{e3}^{}|^2 + |U_{e2}^{}|^2 \alpha -
\left(\beta - \gamma\right)\left(1 - \frac{\kappa_{11}^{}}{2}\right)^2 +
\gamma \left(|\kappa_{21}^{}|^2 + |\kappa_{31}^{}|^2\right) \right] \;,
\nonumber\\
Y^{ee}_3 \hspace{-0.17cm} &\simeq& \hspace{-0.17cm}
\frac{\Delta_{31}^{2}}{4 E^2} \left(1 - \frac{\kappa_{11}^{}}{2}\right)^2
\Biggm\{ \left[|U_{e3}^{}|^2 + |U_{e2}^{}|^2 \alpha -
\left( \beta - \gamma\right)\left(1 - \frac{\kappa_{11}^{}}{2}\right)^2
+ \gamma \left(|\kappa_{21}^{}|^2 + |\kappa_{31}^{}|^2\right)\right]^2
\nonumber\\
&&\hspace{-0.17cm}
+ \bigg|  U_{e3}^{*} U_{\mu 3}^{} + \alpha U_{e2}^{*} U_{\mu 2}^{}
-\gamma \left[\kappa_{21}^{}
\left(1-\frac{\kappa_{22}^{}}{2}\right) -\kappa_{31}^{} \kappa_{32}^*\right]
\bigg|^2 \nonumber\\
&&\hspace{-0.17cm}
+ \bigg| U_{e3}^{*} U_{\tau 3}  +
\alpha U_{e2}^{*} U_{\tau 2} -\gamma \kappa_{31}^{} \left(1 -
\frac{\kappa_{33}^{}}{2}\right) \bigg|^2
\Biggm\} \;,
\end{eqnarray}
where each $Y^{ee}_i$ has a factor $(1-\kappa_{11}^{}/2)^2$. Moreover,
we obtain
\begin{eqnarray}
N_{11}^{} \hspace{-0.17cm} &\simeq& \hspace{-0.17cm}
1 + \frac{1}{\alpha} \left[\left(1 + \alpha\right)\left(\gamma - \beta |U_{e1}^{}|^2\right)
+\frac{\xi_1^{}}{2} \right]+ \frac{1}{\alpha^2}
\left[\beta^2 |U_{e1}^{}|^2 \left(|U_{e1}^{}|^2 - 2 |U_{e2}^{}|^2\right)
\right. \nonumber\\
&& \hspace{-0.17cm}\left.
- \beta\gamma \left(|U_{e1}^{}|^2 - |U_{e2}^{}|^2\right) - \frac{\xi_3^{}}{2}
\right] \;,\nonumber\\
N_{21}^{} \hspace{-0.17cm} &\simeq& \hspace{-0.17cm}
-\frac{1}{\alpha} \left[\left(1 + \alpha\right)\left(\gamma - \beta |U_{e1}^{}|^2\right)
+\frac{\xi_1^{}}{2}\right] -  \frac{1}{\alpha^2}
\left[\beta^2 |U_{e1}^{}|^2 \left(|U_{e1}^{}|^2 - 2 |U_{e2}^{}|^2\right)
\right. \nonumber\\
&& \hspace{-0.17cm}\left.
-\beta\gamma \left(|U_{e1}^{}|^2 - |U_{e2}^{}|^2\right) - \frac{\xi_3^{}}{2}
\right] \;, \nonumber \\
N_{31}^{} \hspace{-0.17cm} &\simeq& \hspace{-0.17cm}
0\;;
\end{eqnarray}
\begin{eqnarray}
N_{12}^{} \hspace{-0.17cm} &\simeq& \hspace{-0.17cm}
\frac{2 E}{\Delta_{31}^{}}\bigg\{-1 + \beta(|U_{e1}^{}|^2 -
|U_{e3}^{}|^2) - \frac{1}{\alpha} \left[1 + 2
\left(\gamma - \beta |U_{e1}^{}|^2\right) +\xi_1^{}\right] +
\frac{1}{\alpha^2}\left[\beta \left(1 + 2 \gamma\right)
\left(|U_{e1}^{}|^2 \right.\right.
\nonumber\\
&&\hspace{-0.17cm}\left.- |U_{e2}^{}|^2) - 2 \beta^2 |U_{e1}^{}|^2
\left(|U_{e1}^{}|^2 - 2 |U_{e2}^{}|^2\right) + \xi_3^{}\right] -
\frac{1}{\alpha^3} \left[\beta^2\left(|U_{e1}^{}|^4
-4 |U_{e1}^{}|^2 |U_{e2}^{}|^2 + |U_{e2}^{}|^4\right)\right.
\nonumber\\
&&\hspace{-0.17cm} \left.- \xi_3^{}\right]
+ \frac{\beta \left(|U_{e1}^{}|^2 -
|U_{e2}^{}|^2\right)}{\alpha^4} \left[\beta^2
\left(|U_{e1}^{}|^4 -6|U_{e1}^{}|^2 |U_{e2}^{}|^2 +|U_{e2}^{}|^4 \right) -
2\xi_3^{}\right]\bigg\} \;,
\nonumber\\
N_{22}^{} \hspace{-0.17cm} &\simeq& \hspace{-0.17cm}
\frac{2 E}{\Delta_{31}^{}}\bigg\{1 - \beta\left(1+|U_{e1}^{}|^2 -
2|U_{e3}^{}|^2\right) + 2 \gamma +\alpha +\alpha^2
+ \frac{1}{\alpha} \left[1 + 2
\left(\gamma - \beta |U_{e1}^{}|^2\right) +\xi_1^{}\right]
\nonumber\\
&&\hspace{-0.17cm}-\frac{1}{\alpha^2}
\left[\beta \left(1 + 2 \gamma\right)\left(|U_{e1}^{}|^2-
|U_{e2}^{}|^2\right) - 2 \beta^2 |U_{e1}^{}|^2
\left(|U_{e1}^{}|^2 - 2 |U_{e2}^{}|^2\right) + \xi_3^{}\right]
\nonumber\\
&&+ \frac{1}{\alpha^3} \left[\beta^2\left(
|U_{e1}^{}|^4 + |U_{e2}^{}|^4\hspace{-0.17cm}
-4 |U_{e1}^{}|^2 |U_{e2}^{}|^2\right)- \xi_3^{}\right]
- \frac{\beta \left(|U_{e1}^{}|^2 -
|U_{e2}^{}|^2\right)}{\alpha^4} \left[\beta^2
\left(|U_{e1}^{}|^4 \right.\right.
\nonumber\\ && \left.\left.
-6|U_{e1}^{}|^2 |U_{e2}^{}|^2 +|U_{e2}^{}|^4 \right) -
2\xi_3^{}\right]\bigg\} \;,
\nonumber\\
N_{32}^{} \hspace{-0.17cm} &\simeq& \hspace{-0.17cm}
-\frac{2 E}{\Delta_{31}^{}}\left[\alpha  +\alpha^2-
\beta \left(1- |U_{e3}^{}|^2\right)+ 2\gamma\right] \;;
\end{eqnarray}
and
\begin{eqnarray}
N_{13}^{} \hspace{-0.17cm} &\simeq& \hspace{-0.17cm}
\frac{4 E^2}{\Delta_{31}^{2}} \bigg\{\frac{1}{\alpha}
\left[1 - \beta \left(|U_{e1}^{}|^2-|U_{e3}^{}|^2\right) \
+\frac{\xi_1^{}}{2} + \xi_2^{}\right] - \frac{1}{\alpha^2}\left[
\beta \left(|U_{e1}^{}|^2 - |U_{e2}^{}|^2\right) - \beta^2
\left(|U_{e1}^{}|^4 \right.\right.
\nonumber\\
&& \left.\left. \hspace{-0.17cm}
- 2|U_{e1}^{}|^2 |U_{e2}^{}|^2 -
|U_{e1}^{}|^2 |U_{e3}^{}|^2 + |U_{e2}^{}|^2 |U_{e3}^{}|^2\right)
+ \frac{\xi_3^{}}{2}\right] + \frac{1}{\alpha^3} \left[
\beta^2 \left(|U_{e1}^{}|^4 - 4 |U_{e1}^{}|^2 |U_{e2}^{}|^2
\right.\right.\nonumber\\
&&\left.\left. \hspace{-0.17cm} + |U_{e2}^{}|^4\right)
- \xi_3^{}\right] - \frac{\beta\left(|U_{e1}^{}|^2 -|U_{e2}^{}|^2\right)}
{\alpha^4}\left[\beta^2 \left(|U_{e1}^{}|^4 - 6 |U_{e1}^{}|^2
|U_{e2}^{}|^2 + |U_{e2}^{}|^4\right) - 2 \xi_3^{}\right]
\bigg\} \;, \nonumber\\
N_{23}^{} \hspace{-0.17cm} &\simeq& \hspace{-0.17cm}
\frac{4 E^2}{\Delta_{31}^{2}} \bigg\{-1-\alpha \left(1 + \alpha\right) +
\beta \left(1-3 |U_{e3}^{}|^2\right) - \frac{1}{\alpha}
\left[1 - \beta \left(|U_{e1}^{}|^2-|U_{e3}^{}|^2\right) \
+\frac{\xi_1^{}}{2} + \xi_2^{}\right]
\nonumber\\
&&\hspace{-0.17cm}
+ \frac{1}{\alpha^2}\left[\beta \left(|U_{e1}^{}|^2
- |U_{e2}^{}|^2\right) - \beta^2\left(|U_{e1}^{}|^4
- 2|U_{e1}^{}|^2 |U_{e2}^{}|^2 -
|U_{e1}^{}|^2 |U_{e3}^{}|^2 + |U_{e2}^{}|^2 |U_{e3}^{}|^2\right)
+ \frac{\xi_3^{}}{2}\right] \nonumber\\
&& \hspace{-0.17cm}
- \frac{1}{\alpha^3} \left[
\beta^2 \left(|U_{e1}^{}|^4 - 4 |U_{e1}^{}|^2 |U_{e2}^{}|^2
+ |U_{e2}^{}|^4\right) - \xi_3^{}\right] + \frac{\beta\left(|U_{e1}^{}|^2
-|U_{e2}^{}|^2\right)}{\alpha^4}\left[\beta^2 \left(|U_{e1}^{}|^4
\right.\right.\nonumber\\
&&\left.\left.\hspace{-0.17cm}- 6 |U_{e1}^{}|^2
|U_{e2}^{}|^2 + |U_{e2}^{}|^4\right) - 2 \xi_3^{}\right]
\bigg\} \;,
\nonumber\\
N_{33}^{} \hspace{-0.17cm} &\simeq& \hspace{-0.17cm}
\frac{4 E^2}{\Delta_{31}^{2}} \left[1 + \alpha + \alpha^2 -
\beta \left(1 - 3 |U_{e3}^{}|^2\right)\right]\;.
\end{eqnarray}
It is clear that $N_{11}^{} + N_{21}^{} + N_{31}^{}=1$,
$N_{12}^{} + N_{22}^{} + N_{32}^{}=0$ and $N_{13}^{} + N_{23}^{}
+ N_{33}^{}=0$ hold. These three relations are exactly valid, as
one can see from Eq. (7).

\end{document}